\documentclass[superscriptaddress]{book}

\usepackage{graphicx,graphics,epsfig,subfigure,times,bm,bbm,amssymb,amsmath,amsfonts,amsthm,mathrsfs,MnSymbol}
\usepackage[T1]{fontenc}
\usepackage{comment}

\usepackage[matrix,frame,arrow]{xypic}
\usepackage[pdfstartview=FitH]{hyperref}
\usepackage{pifont}

\usepackage{times}
\usepackage{braket}
\usepackage{multirow}
\usepackage[normalem]{ulem}

\usepackage{subfigure}
\usepackage{overpic}
\usepackage{fancyhdr}

\newcommand{\be}{\begin{equation}}
\newcommand{\ee}{\end{equation}}
\newcommand{\ba}{\begin{eqnarray}}
\newcommand{\ea}{\end{eqnarray}}
\newcommand\tr{{\mbox{Tr\,}}}
\newcommand{\ignore}[1]{}
\newcommand{\bea}{\begin{eqnarray}}
\newcommand{\eea}{\end{eqnarray}}

\def\ii{\mathrm{i}}

\newcommand{\eu}{{\rm e}}
\newcommand{\de}{{\displaystyle\rm\mathstrut d}}

\usepackage[margin=1in]{geometry}
\usepackage[onehalfspacing]{setspace}

\newcommand{\chapabstract}[1]{
	\begin{quote}
		\singlespacing\small
		\rule{14cm}{1pt}
		#1
		\vskip-4mm
		\rule{14cm}{1pt}
\end{quote}}

\makeatletter
\newcommand{\chapterauthor}[1]{%
	{\parindent0pt\vspace*{-25pt}%
		\linespread{1.1}\large\scshape#1%
		\par\nobreak\vspace*{35pt}}
	\@afterheading%
}
\makeatother

\usepackage{etoolbox}
\makeatletter
\patchcmd{\thebibliography}{%
	\chapter*{\bibname}\@mkboth{\MakeUppercase\bibname}{\MakeUppercase\bibname}}{%
	\section{Bibliography}}{}{}
\makeatother

\begin{document}

\chapter*{Local Convertibility in Quantum Spin Systems} 

\thispagestyle{fancy}
\rhead{RBI-ThPhys-2022-1}

\chapterauthor{Luigi Amico$^1$, Vladimir Korepin$^2$, Alioscia Hamma$^3$, Salvatore Marco Giampaolo$^4$, \& Fabio Franchini$^{4\dagger}$ \\
\\
$^1$ Quantum Research Centre,Technology Innovation Institute, Abu Dhabi, UAE \\
Centre for Quantum Technologies, National University of Singapore, 3 Science Drive 2, Singapore 117543 \\
On leave from Dipartimento di Fisica e Astronomia, Via S. Sofia 64, 95127 Catania, Italy\\
LANEF 'Chaire d’excellence’, Universit\`e Grenoble-Alpes \& CNRS, F-38000, Grenoble,France \\
$^2$ C. N. Yang Institute for Theoretical Physics, Stony Brook University, NY 11794, USA \\
$^3$ Physics Department, University of Massachusetts Boston, 02125, USA \\
$^4$ Ru\dj er Bo\v{s}kovi\'c Institute, Bijeni\v{c}ka cesta  54,  10000 Zagreb, Croatia \\
$^\dagger$ email: fabio@irb.hr}

\chapabstract{\\
Local Convertibility refers to the possibility of transforming a given state into a target one, just by means of LOCC with respect to a given bipartition of the system and it is possible if and only if all the R\'{e}nyi-entropies of the initial state are smaller than those of the target state. We apply this concept to adiabatic evolutions and ask whether they can be rendered through LOCC in the sense above. We argue that a lack of differential local convertibility (dLC) signals a higher computational power of the system's quantum phase, which is also usually connected with the existence of long-range entanglement, topological order, or edge-states. Remarkably, dLC can detect these global properties already by considering small subsystems. Moreover, we connect dLC to spontaneous symmetry breaking by arguing that states with finite order parameters must be the most classical ones and thus be locally convertible.
}

\begingroup
\let\cleardoublepage\clearpage
\tableofcontents
\endgroup

\vskip1cm

\pagestyle{headings}

\setcounter{chapter}{1}

\section[Introduction]{Introduction}
\markboth{Introduction}{Introduction}

Entanglement is one of the primary resources for quantum technology\cite{entanglement1,entanglement2,iontraps, coldatoms, superconductivity, photonics, nmr}, as it encodes the possibility of storing a large amount of information on a registry on one side and to change a state even in parts that are not directly manipulated\cite{feynman1982,lloyd1996}. However, it is also clear that is not the mere amount of entanglement which is important, but rather how it is distributed and how effectively it can be used\cite{algorithm, gottesman, vidal2003a, vandernest2007, Briegel08, winter, eisert, Briegel09, vandernest2013, Raussendorf13}.

The quantum adiabatic algorithm provides a good paradigm to test the properties of a given phase of matter, as it has been proven to provide a universal platform for quantum computation \cite{adiabatic}. 
It is based on the idea that starting from a simple system with a known ground state, through a suitable adiabatic evolution the final ground state can encode the result of a computation or the state of a quantum system one aims at simulating. Since the closing of the gap forces a dramatic slowdown in the rate at which the Hamiltonian can be changed and the system expected to remain in its instantaneous ground state, crossing a phase transition through the evolution typically impairs the efficacy of an adiabatic algorithm. Hence, for this algorithm to provide a nontrivial advantage the choice of the initial quantum phase is crucial. 

A criterion for this choice is provided by the concept of {\it differential Local Convertibility} (dLC), which addresses the question of whether it is possible to reproduce the adiabatic evolution of a bipartite quantum system through Local Operations and Classical Communications (LOCCs) \cite{Nielsen99, Jonathan-Plenio}.
Namely, upon partitioning a many-body system into two blocks $A$ and $B$, one questions whether the response of the ground state $|0\rangle$ to an external perturbation can be rendered through LOCC {\it restricted to $A$ and $B$ individually}? If affirmative, the ground state can be moved around within a given quantum phase by LOCC. If negative the adiabatic evolution induced by the perturbation cannot be captured classically (due to the long-range coherence present in the system).
Quantitatively, it accounts for the response of the  R\'{e}nyi entropy
\be
S_\alpha \doteq { {1}\over{1-\alpha} }  \log \tr \rho_A^\alpha 
\ee 
to the changing of a control parameter $g$ in the Hamiltonian. Here, $\rho_A \doteq \tr_B |0\rangle \langle 0|$ is the reduced density matrix of the block $A$ and $\alpha$ is a free parameter which tunes different entanglement measures \cite{NielsenChuang}. For instance, while low $\alpha$'s weight more evenly all eigenvalues of $\rho_A$, higher values of $\alpha$ enhance the role of the larger eigenvalues. If all the R\`enyi entropies decrease along a given path, this evolution can also be rendered through LOCC (at least concerning the chosen partition of $A$ and $B$) \cite{Turgut, Klimesh} and thus the adiabatic algorithm cannot provide a significant improvement. This observation was at the heart of  \cite{Cui-Gu, Franchini2014}.

Na\"ively, one expects entanglement to always grow moving toward a phase transition. In particular, $S_0 =\log R$, where $R$ is the Schmidt rank of the state, i.e. the number of nonzero eigenvalues of $\rho_A$ (while $S_1$ is the von Neumann entropy measuring the entanglement entropy for the subsystem $A$).  Generically, $R$ and thus $S_0$ increase with $\xi$  because more degrees of freedom get entangled by increasing the range of correlations. Nonetheless, the study of Local Convertibility shows that the picture is more complex when higher $\alpha$ entropies are considered. Certain systems can support other forms of entanglement not captured by local correlation, which are referred to as long-range entanglement (LRE) \cite{Wen}. If the latter would decrease approaching a phase transition, the competition between the different forms of entanglement could be detected as a lack of dLC in all directions.

A simple form of LRE is connected to the existence of edge states at distant boundaries of a system \cite{hasan2010,mourik2012}.
Furthermore, these edge states are usually the reflection of the existence of some sort of topological order. Interestingly, upon partitioning a system into two, new pairs of edge states are created also at the boundaries of the partitions. While for thermodynamic systems the LRE connecting the edge states is maximal, when the size of one of the partitions becomes comparable with the correlation length, the edge state can undergo a process of recombination which reduces their LRE. Through this mechanism, higher R\`enyi entropies can decrease as the correlation length increases thus destroying dLC in any directions of the adiabatic evolution.
The behavior of dLC in relation to quantum phase transitions and at criticality has been analyzed, respectively, in \cite{braganca2014} and \cite{huai2014}.

In addition to shedding new light on the role of edge states in providing a key advantage in a universal quantum computational platform, dLC also proves useful in identifying phases characterized by LRE. LRE is defined as that entanglement that cannot be destroyed by reducing the state to a trivial (factorized) one through a finite depth quantum circuit \cite{Wen}. As such, it is usually revealed through nonlocal string order parameters whose lengths exceed the order of usual correlations. Remarkably, dLC can detect LRE using partitions of the order of the correlation length, thus providing a somewhat local probe of an elusive long-range property \cite{Cui-Amico, Hamma-Cincio, Santra2014, dai2015, tzneg2016}.
We will analyze a few examples of models displaying topological order in light of their local convertibility. Topological phases have attracted a lot of attention for their ability to defy the Ginzburg--Landau paradigm by not having any finite local order parameter, but rather a topological one \cite{Wen-1, Wen-protected}.

However, dLC has also been connected to the usual spontaneous symmetry breaking mechanism responsible for the ensuing of local order \cite{Cianciaruso}. 
In particular, on the whole, the complete understanding of the physical mechanism that selects the symmetry-breaking ground states in the thermodynamic limit remains an open problem~\cite{Bratteli2012, Arodz2012}.
In complete analogy with the case of classical phase transitions driven by temperature, the common explanation of this phenomenon invokes the unavoidable presence of some local, however small, perturbing external field that selects one of the maximally symmetry-breaking ground states (MSBGSs) among all the elements of the quantum ground space \cite{S2000}. Crucially, in this type of reasoning, it is assumed that the MSBGSs are the most classical ones and thus the ones that are selected in real-world situations, under the effect of decoherence that quickly destroys macroscopic coherent superpositions.

At first glance, this notion appears to be obvious. For instance, in the paradigmatic case of the quantum Ising model, the ground space of the ferromagnetic phase at zero transverse fields $h$ is spanned by two orthogonal product states $|0\rangle^{\otimes N}$ and $|1\rangle^{\otimes N}$ which are in the same class of pointer states of the typical decoherence argument, while the symmetric states $\Psi_{\pm} = 1/\sqrt{2}(|0\rangle^{\otimes N} \pm |1\rangle^{\otimes N})$ realize macroscopic coherent superpositions (Schrodinger cats) that are not stable under decoherence~\cite{Zurek2003,vanWezel2008}. Therefore, at zero transverse fields $h$,
the situation is very clear: the only stable states are those that maximally break the symmetry of the Hamiltonian and at the same time
those that feature vanishing macroscopic total correlations, including entanglement, between spatially separated regions.

On the other hand, as we turn on the external field $h$, we have a whole range of values, below the critical field $h=h_c$, where it remains a finite
magnetic order associated with spontaneous symmetry breaking~\cite{mcoy}, which implies an application of the decoherence argument within the entire, globally ordered phase. This means that, again, the only stable states are those that maximally break the Hamiltonian symmetry.
However, now the symmetry-breaking states are entangled, and their mixed-state reductions on arbitrary subsystems possess in general
nonvanishing pairwise entanglement~\cite{Osborne2002, Osterloh2002, entanglement1}, as well as pairwise quantum~\cite{TRHA2013, Tomasello2012, Campbell2013}
and classical correlations~\cite{mcoy}. It is thus now unclear if and in what sense the MSBGSs are the most classical among all quantum
ground states. 

Below we will provide a general conjecture on the nature of ordered quantum
phases and the origin of spontaneous symmetry breaking, by comparing various quantifiers of local and global quantum correlations in symmetry-breaking and symmetry-preserving quantum ground states. We will first compare measures of local, pairwise quantum correlations and show that in symmetry-preserving ground states the two-body entanglement captures only a modest portion of the local, two-body quantum correlations, while in maximally symmetry-breaking ground states it accounts for the largest contribution. Next, we will introduce proper criteria and quantifiers of the degree of classicality of quantum states for their global contents of macroscopic entanglement and quantum correlations. Finally, we will show that, within the quantum ground space corresponding to macroscopically ordered phases with nonvanishing local order parameters, the MSBGSs are the most classical ground states in the sense that they are the only quantum ground states that satisfy the following two criteria for each set of Hamiltonian parameters consistent with an ordered quantum phase in the thermodynamic limit:
\begin{itemize}
	\item {\em Local convertibility} -- All global ground states are convertible into MSBGSs applying only local operations and classical communication (LOCC transformations), while the reverse transformation is impossible.
	\item {\em Entanglement distribution} -- The MSBGSs are the only global ground states that minimize the residual tangle between a dynamical variable and the remainder of a macroscopic quantum system. Stated otherwise, the MSBGSs are the only ground states that satisfy the monogamy inequality -- a strong constraint, with no classical counterpart, on the shared bipartite entanglement between all components of a macroscopic quantum system -- at its minimum among all other possible ground states and thus minimize the macroscopic multipartite entanglement as measured by the residual tangle.
\end{itemize}
Verification of these two features amounts to proving that the mechanism of spontaneous symmetry breaking selects the most classical ground states associated with globally ordered phases of quantum matter with nonvanishing local order parameters.

This chapter is organized as follows: in Sects. \ref{convert_clusterising}, \ref{convert_lambda-D}, and \ref{convert_pertToric} we will analyze the dLC of the Cluster-Ising chain, of the $\lambda-D$ model and of the two-dimensional toric code with different perturbative terms, to show how all these models, characterized by different types of topological order, are not locally convertible. In Sect. \ref{QIsingSec} we will use the paradigmatic example of the Quantum Ising Chain to elucidate the role of edge states in dLC and thus to provide a picture of how LRE prevents convertibility and why small partitions can detect it. In Sect. \ref{classicalsec} we detail the conjecture on the characterization of MSBGS as the most classical ones. Finally, we draw some conclusions in Sect. \ref{conclusionSec}.

\section{The Cluster-Ising Model}
\label{convert_clusterising}
\markboth{The cluster-Ising model}{The cluster-Ising model}

\begin{figure}[h]
	\includegraphics[width=\columnwidth]{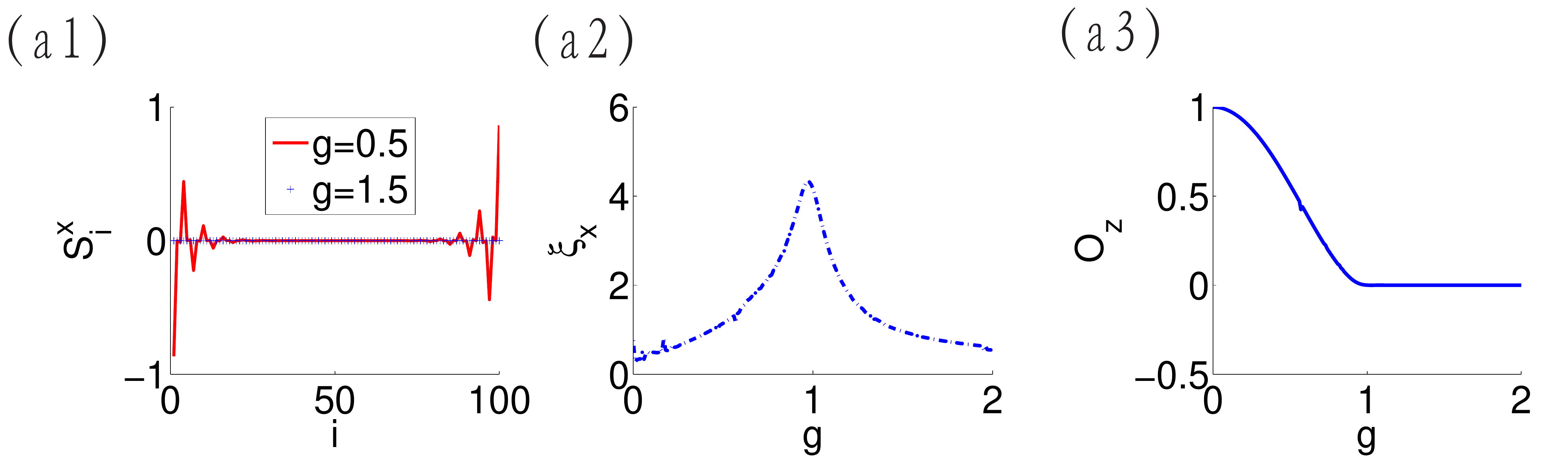}
	\caption{The edge state, correlation length, and the string order parameter of the Cluster-Ising model.  $(a1)$ shows that there is an edge state in the cluster phase, whereas there is no edge state in Ising antiferromagnetic phase. $(a2)$ shows  the correlation length of $\langle \sigma_n \sigma_{n+3}\rangle-\langle \sigma_n \rangle\langle\sigma_{n+3}\rangle$ displaying a critical behavior. $(a3)$ is the string order parameter ${\cal O}_z=(-)^{N-2}\langle \sigma_1^y\prod_{j=1}^{N-1} \sigma_j^z \sigma_N^y\rangle$.}
	\label{stat_cluster}
\end{figure}

The first Hamiltonian we consider is
\begin{equation}
	H(g)=-\sum_{j=1}^N \sigma_{j-1}^x \sigma_j^z \sigma_{j+1}^x + g \sum_{j=1}^N \sigma_j^y \sigma_{j+1}^y,
	\label{eq:ham_cluster}
\end{equation}
where $\sigma_i^{\alpha}$, $\alpha=x,y,z$, are the Pauli matrices and, except otherwise stated, we take open  boundary conditions $\sigma^{\alpha}_{N+1}= \sigma^{\alpha}_{0}=0$.
The phase diagram of (\ref{eq:ham_cluster}) has been investigated in \cite{Son11,smacchia}. For large $g$, the system is an Ising antiferromagnet with a finite local order parameter. For $g=0$, the ground state is a cluster state. It results that the correlation pattern characterizing the cluster state  is robust up to a critical value of the control parameter, meaningfully defining a ``cluster phase" with vanishing order parameter and string order\cite{Son11,smacchia}. Without symmetry, the cluster phase is a (non-topological) quantum spin liquid since there is a gap and no symmetry is spontaneously broken. Protected by a $Z_2 \times Z_2$ symmetry, the cluster phase is characterized by a topological fourfold ground state degeneracy, reflecting the existence of edge states and fanning out from $g=0$ where $4$ Majorana fermions are left free at the ends of the chain \cite{Son11,Son12}.  In the DMRG, we resolve the ground state degeneracy,  by adding a small perturbation $\sigma_1^x\sigma^z_2\pm\sigma^z_{N-1}\sigma_N^x$ to the Hamiltonian. The Cluster and Ising phases are separated by a continuous quantum phase transition with central charge $c=3/2$. Let us also note that the Hamiltonian (\ref{eq:ham_cluster}) can be mapped to three decoupled Ising chains \cite{Son11,smacchia}.

\begin{figure}
	\begin{minipage}[c]{0.62\textwidth}
		\includegraphics[width=\textwidth]{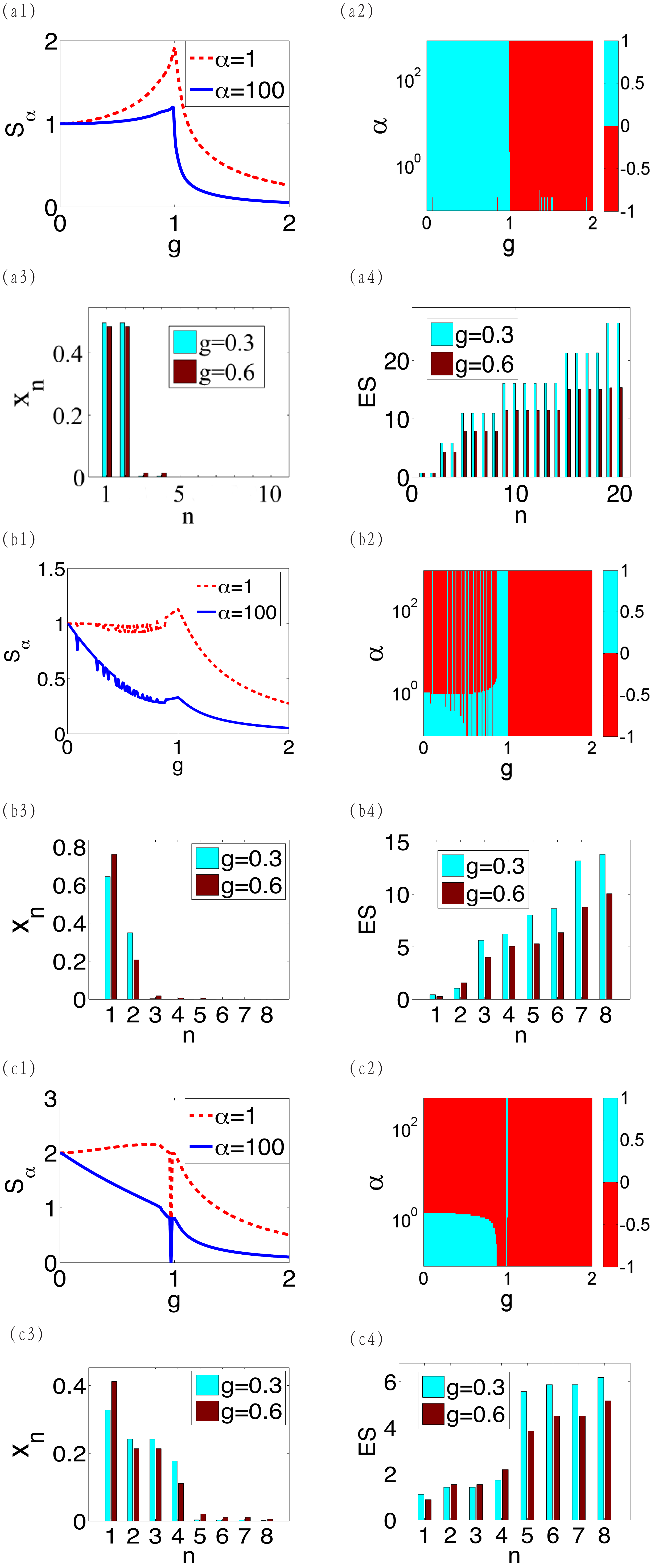}
	\end{minipage}\hfill
	\begin{minipage}[c]{0.35\textwidth}
		\caption{The local convertibility  and the entanglement spectrum  of the Cluster-Ising model Eq.(\ref{eq:ham_cluster}).
			We characterize the differential local convertibility in terms of the slopes of the R\'{e}nyi entropies.
			Panel (a) is for bipartition $A|A$, $A=50$. There is differential local convertibility throughout the two different phases because,  for fixed $g$, $\partial_g S_\alpha$ does not change sign with $\alpha$.   Panel (b) is for bipartition $A|B$, $A=3$, $B=97$. Panel (c) is for $A|B|C$, being one block $A\cup C$ with $A=48$, $C=49$, and $B=3$.  In all of these cases, $\partial_g S_\alpha$ changes sign.
			Panels (a3,b3,c3) and (a4,b4,c4) display respectively the reduced density matrix eigenvalues $x_n$ and the entanglement spectrum  $ES\doteq \left \{ -\log x_n\right \}$. In convertible phases, we observe that the change in the largest eigenvalues is ``faster'' than the rate at which the smallest eigenvalues are populated. In contrast the  non-differential local convertibility arises because the sharpening of the first part of the spectrum is over-compensated by the increasing of the smallest $x_n$.} \label{convert_cluster}
	\end{minipage}
\end{figure}


Through a Jordan--Wigner transformation
$
\sigma_k^{+}=c_k^{\dag}\prod_{j<k}\sigma_j^z$, 
$\sigma_k^{-}=c_k\prod_{j<k}\sigma_j^z$, 
$\sigma_k^z=2c_k^{\dag}c_k-1$, 
the Hamiltonian of the Cluster-Ising model can be written as
\begin{eqnarray}
	H(g) 
	=-i \sum_k\Big[f_k^{(2)}f_{k+2}^{(1)}-gf_k^{(1)}f_{k+1}^{(2)}\Big].
\end{eqnarray}
where $f_k^{(1)}=c_k+c_k^{\dag}$ and  $f_k^{(2)}=-i(c_k-c_k^{\dag})$
are two types of Majorana fermion operators.
Although no local order parameters exist to characterize the topological phase, the topological order in the Cluster-Ising model can be detected, see Fig. \ref{stat_cluster}, by the edge states (a1) and string order parameters (a3).

\begin{figure}
		\begin{minipage}[c]{0.57\textwidth}
		\includegraphics[width=\textwidth]{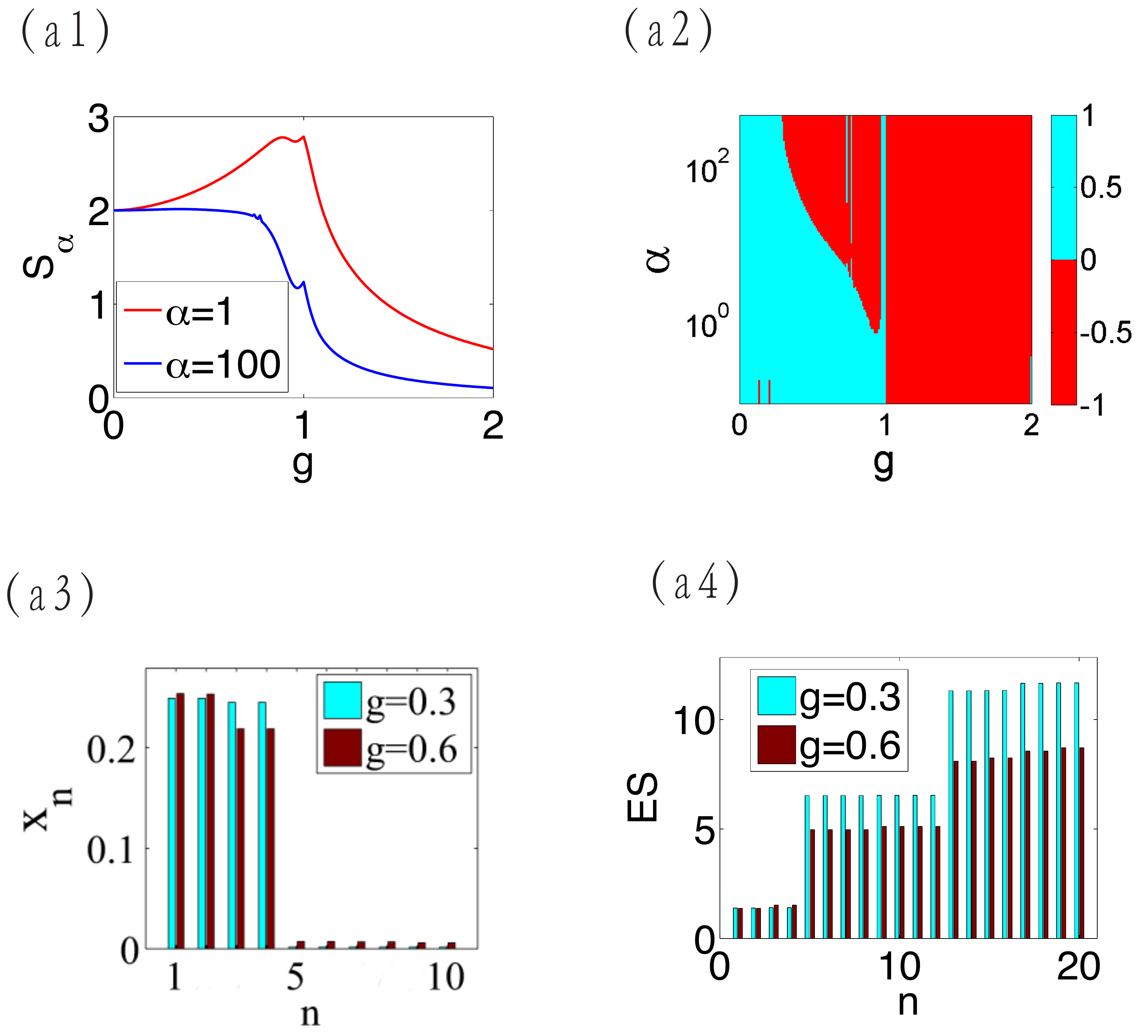}
	\end{minipage}\hfill
	\begin{minipage}[c]{0.4\textwidth}
		\caption{The local convertibility  and the entanglement spectrum  of the Cluster-Ising model with bipartition $45|10|45$. We characterize the differential local convertibility in terms of the slopes of the R\'{e}nyi entropies.
		$\partial_g S_\alpha$ changes sign in the cluster phase.
		$(a3)$  and $(a4)$ display  the largest and the smaller  eigenvalues  of reduced density matrix $x_n$, respectively;  $ES\doteq \left \{ -\log x_n\right \}$. In convertible phases, we observe that the change in the largest eigenvalues is ``faster'' than the rate at which the smallest eigenvalues are populated. In contrast, the  non-differential local convertibility arises because the sharpening of the first part of the spectrum is over-compensated by the increasing of the smallest $x_n$} \label{convert_large_cluster}
	\end{minipage}
\end{figure}

We find that the symmetric partition $A|A$ displays local convertibility, Fig. \ref{convert_cluster}: $(a1), (a2)$. This is indeed a fine-tuned phenomenon since the cluster phase results nonlocally convertible,  for a generic block of spins, both of the type $A|B$ and the  $B|A|B$, Fig. \ref{convert_cluster}. We remark that such a property holds even for size region $A$ smaller than the correlation length.  Indeed, the entanglement spectrum is doubly degenerate in all the cluster phases, as far as the size of the blocks $A$ and $B$ is larger than the correlation length, see Fig.\ref{convert_large_cluster}. In contrast, the antiferromagnet is locally convertible, with nondegenerate entanglement spectrum.


\section{The $\lambda-D$ Model}
\label{convert_lambda-D}
\markboth{The $\lambda-D$ model}{The $\lambda-D$ model}

\begin{figure}[h]
	\hspace*{-0.6cm}\includegraphics[width=1.1\columnwidth]{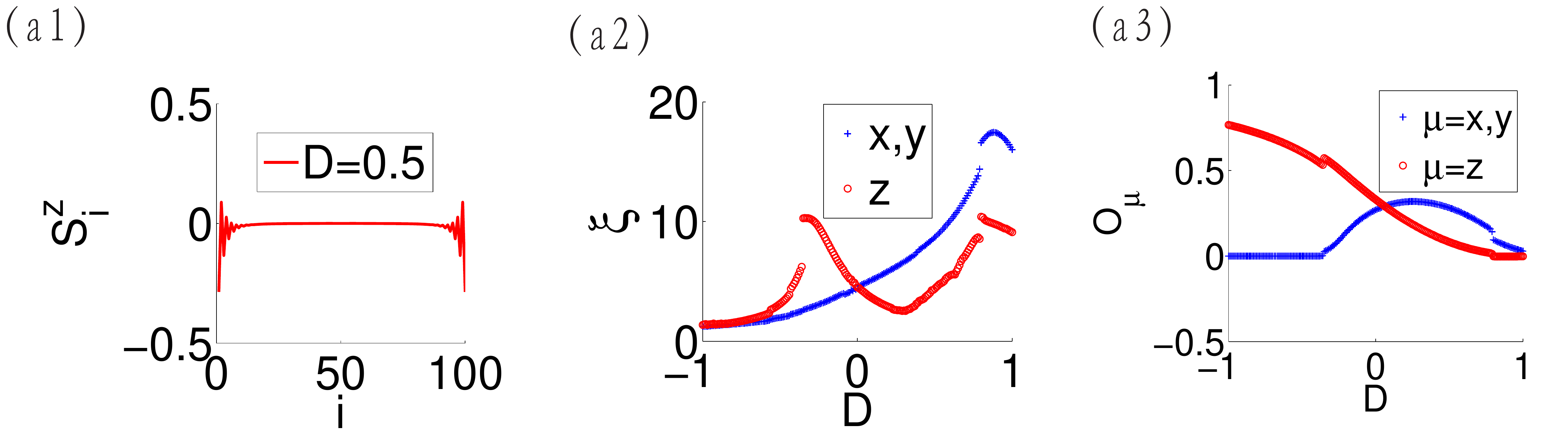}
	\caption{The  edge states, correlation lengths and string order parameters of the $\lambda-D$ model. The sweep (1) through the $\lambda-D$ phase diagram is considered (see text).
		In $(a1)$ we show the Haldane phase edge states; we do not find edge states in the other phases. In  $(a2)$ the string order parameters ${\cal O}_u=(-)^{N-2}\langle S_1^u\prod_{j=1}^{N-1} e^{i\pi S_j^u} S_N^{u}\rangle$. In   $(a3)$ the correlation length of $\langle  S_j^u S_{j+n}^u \rangle - \langle  S_j^u \rangle \langle S_{j+n}^u \rangle$}
	\label{stat_lambdaD}
\end{figure}

In this section, we study the local convertibility of the   $\lambda-D$ model Hamiltonian describing {an interacting spin-$1$ chain} with  a single-ion anisotropy
\begin{eqnarray}
	H=\sum_i[(S_i^xS_{i+1}^x+S_i^yS_{i+1}^y)+\lambda S_i^zS_{i+1}^z +D (S_i^z)^2].
\end{eqnarray}
where $S^u$, $u=\{x,y,z\}$ are spin-$1$ operators: $S^z|\pm\rangle=\pm |\pm \rangle$ and $S^z|0\rangle=0$. The  phase diagram has been investigated by many authors\cite{lambdaDphasediagram,ercolessi,precise_dmrg}.
The Hamiltonian above enjoys several  symmetries,  including  time reversal $S^{x,y,z} \rightarrow -S^{x,y,z} $, parity $S^{x,y} \rightarrow -S^{x,y} $,  $S^{z} \rightarrow S^{z} $ generating  $Z_2\times Z_2$,  and  the  link inversion  symmetry $S^u_j\rightarrow S^u_{-j+1}$.

\begin{figure}
	\begin{minipage}[c]{0.5\textwidth}
		\includegraphics[width=\textwidth]{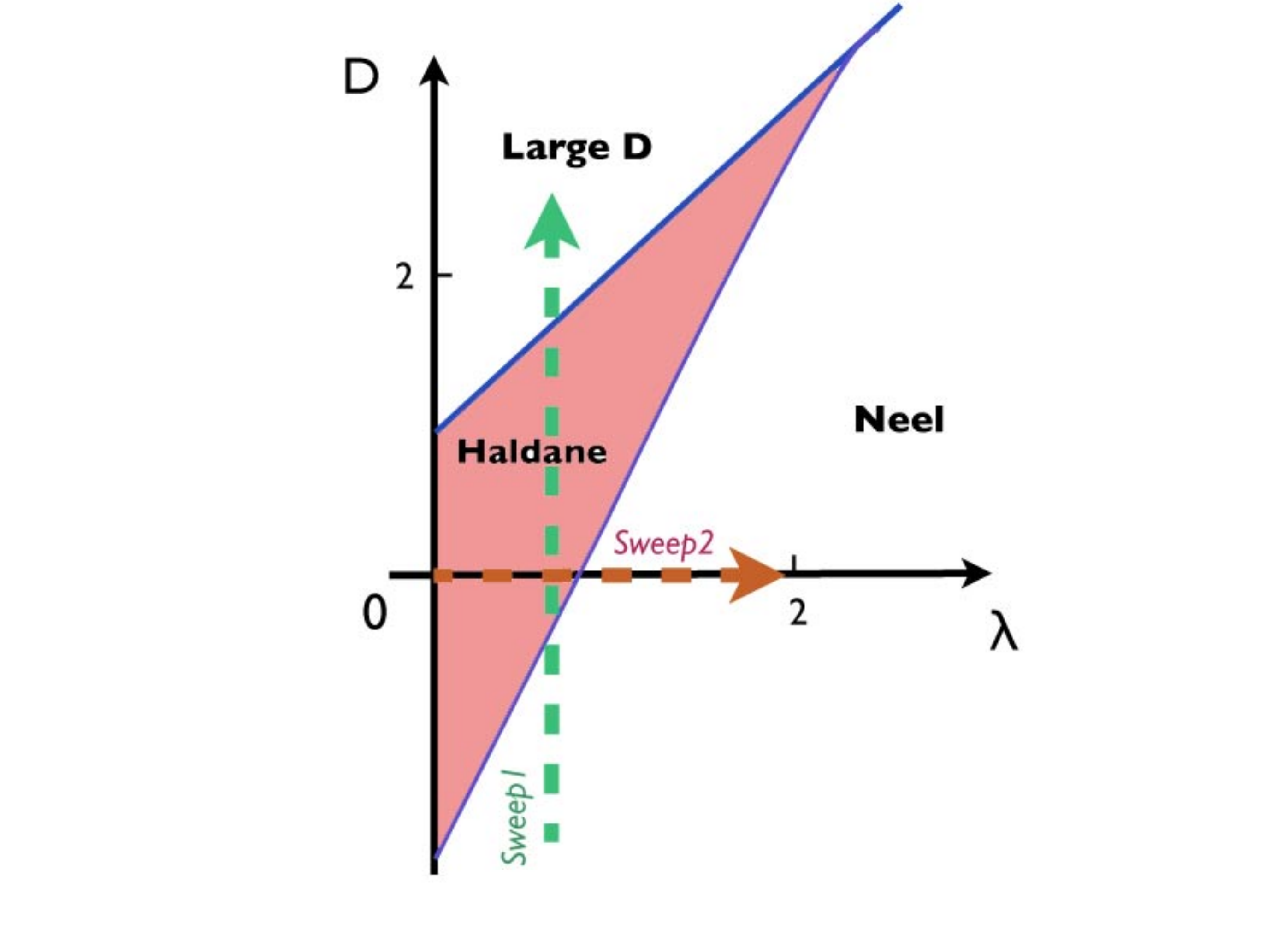}
	\end{minipage}\hfill
	\begin{minipage}[c]{0.45\textwidth}
		\caption{We sweep through  the  phase diagram  in the following two ways: ({\it 1}) fix $\lambda=1$ and change $D$; the Haldane phase is approximately located in the range $-0.4\lesssim D\lesssim 0.8$.  ({\it 2}) Fix $D=0$, varying on $\lambda$; the Haldane phase is located in the range   $0\lesssim \lambda\lesssim 1.1$} \label{sweeps}
	\end{minipage}
\end{figure}

\begin{figure}
	\begin{minipage}[c]{0.6\textwidth}
		\includegraphics[width=\textwidth]{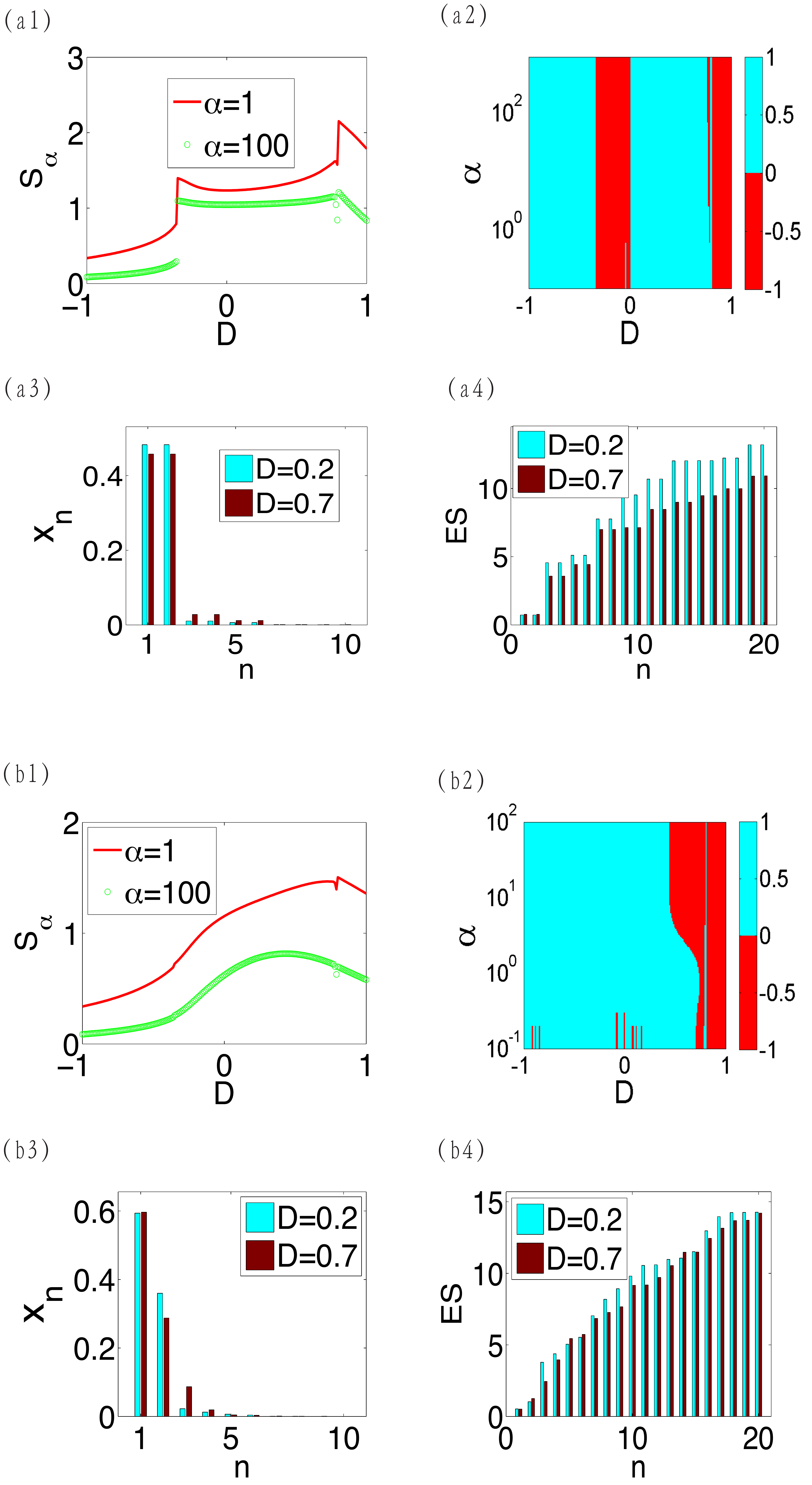}
	\end{minipage}\hfill
	\begin{minipage}[c]{0.37\textwidth}
		\caption{The local convertibility for the partition $A|B$. The sweep (1) through the $\lambda-D$ phase diagram is considered (see also Fig. \ref{sweeps} for the schematic phase diagram). The upper panels display the results for the symmetric case $A|A$.  The bottom panels refer to the antisymmetric case $A=96$, $B=4$. The R\'enyi entropies  are presented in $(a1)$,  $(b1)$. The sign distributions of the entropies derivatives are shown in  $(a2)$, $(b2)$.  The eigenvalues of reduced density matrix $x_n$ and the entanglement spectrum are shown in $(a3)$,$(a4)$, $(b3)$, $(b4)$ as in Fig.\ref{convert_cluster}. The features of differential local convertibility are characterized by the slopes of the R\'enyi entropies and correspond to specific features of the entanglement spectrum as explained in Fig.\ref{convert_cluster}}
		\label{conver_lambdaD_symm}
	\end{minipage}
\end{figure}

We only consider $\lambda > 0$.
For small/large $D$ and fixed $\lambda$, the system is in a polarized state along $|+ \rangle \pm| - \rangle$ or $|0\rangle$, respectively. For large $\lambda$ and fixed $D$,  the  state displays  antiferromagnetic order.
At intermediate $D$ and $\lambda$, the state is a ``diluted antiferromagnet'' with strong quantum fluctuations, defining the  Haldane phase, which cannot be characterized through any local order parameter. With symmetry protection, the topological order in the Haldane phase can be detected by the edge states and string order parameters defined in Fig. \ref{stat_lambdaD} (see \cite{sorensen}). Without symmetry, the ground state is gapped and no symmetry is spontaneously broken, making the Haldane phase a quantum spin liquid. 
In fact, for open boundary conditions (which we apply in the present analysis), the Haldane ground state displays a fourfold degeneracy, which cannot be lifted without breaking the abovementioned symmetry of the  Hamiltonian. This is the core mechanism defining the Haldane phase as a  symmetry-protected topological ordered phase\cite{AKLT-pollmann,AKLT-protected}.

\begin{figure}[h!]
	\begin{minipage}[c]{0.55\textwidth}
		\includegraphics[width=\textwidth]{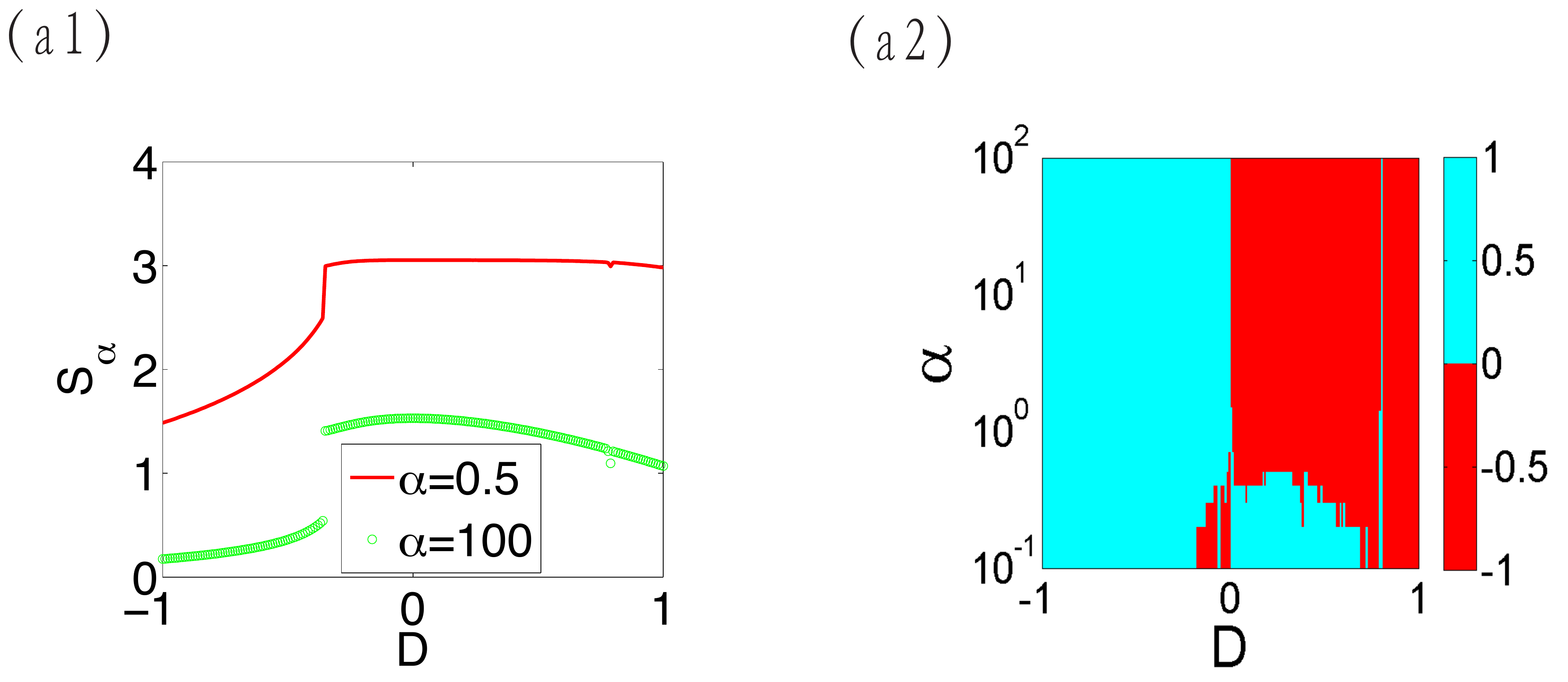}
		\vskip.5cm
		\includegraphics[width=\textwidth]{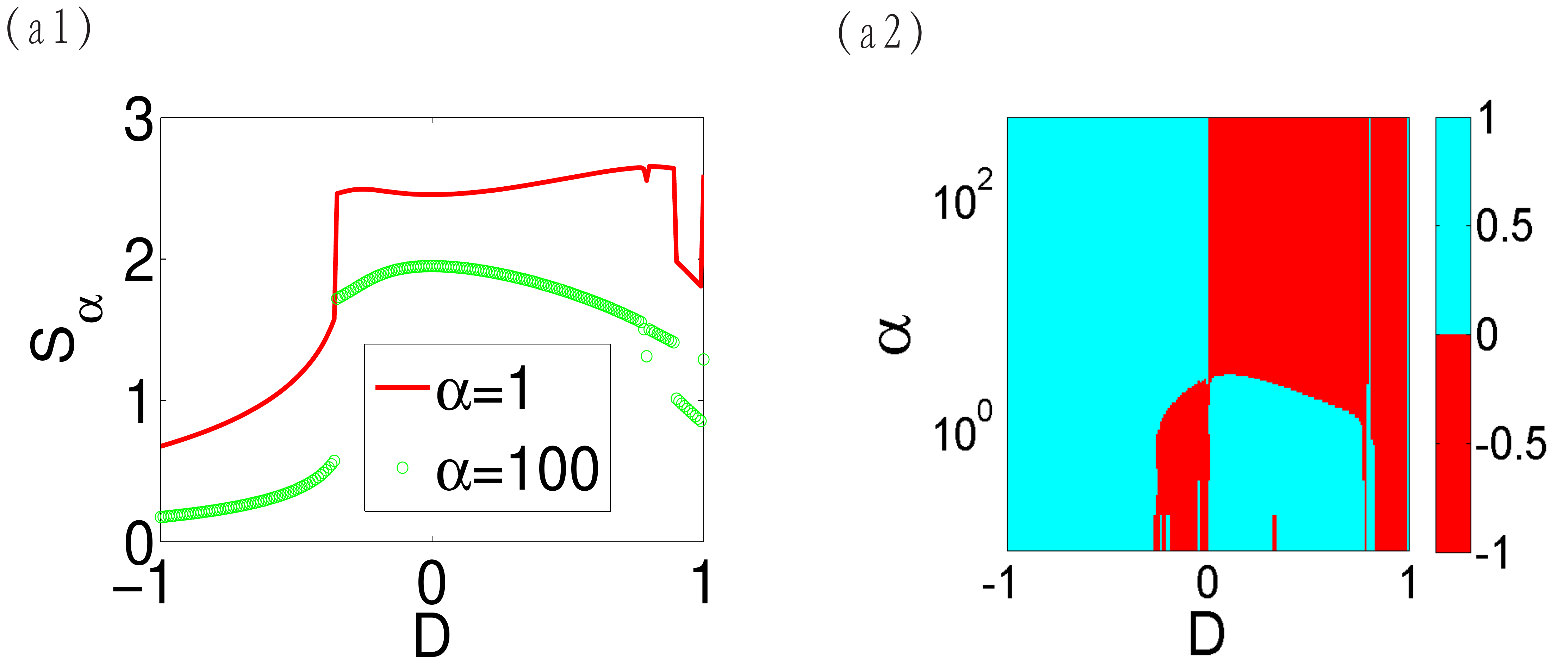}
	\end{minipage}\hfill
	\begin{minipage}[c]{0.42\textwidth}
		\caption{\label{fig_sweep1} 
			Sweep (1) through the $\lambda-D$ model: $\lambda=1$, $D\in \{-1,1\}$. The sign distribution of the derivative of  the R\'{e}nyi entropies $\partial_D S_\alpha$  for partitions $A|B|A$,  $A=48$ and $B=4$ (upper panels) and $A=45$ and $B=10$ (lower panels); both with $N=100$ are presented in $(a2)$. The features of differential local convertibility are characterized by the slopes of the R\'enyi entropies and correspond to specific features of the entanglement spectrum as explained in Fig.\ref{convert_cluster}. The $S_\alpha$'s are presented in $(a1)$ for  $\alpha=0.5,100$ decreasing from top to low.  All such quantities are calculated for  the ground  state in $S_z^{tot}=1$ sector}
		\label{convert_large_lambda-D_1}
	\end{minipage}
\end{figure}
%
%
\begin{figure}[h!]
	\begin{minipage}[c]{0.54\textwidth}
		\includegraphics[width=\textwidth]{PRB88/figure4.pdf}
		\vskip.5cm
		\includegraphics[width=\textwidth]{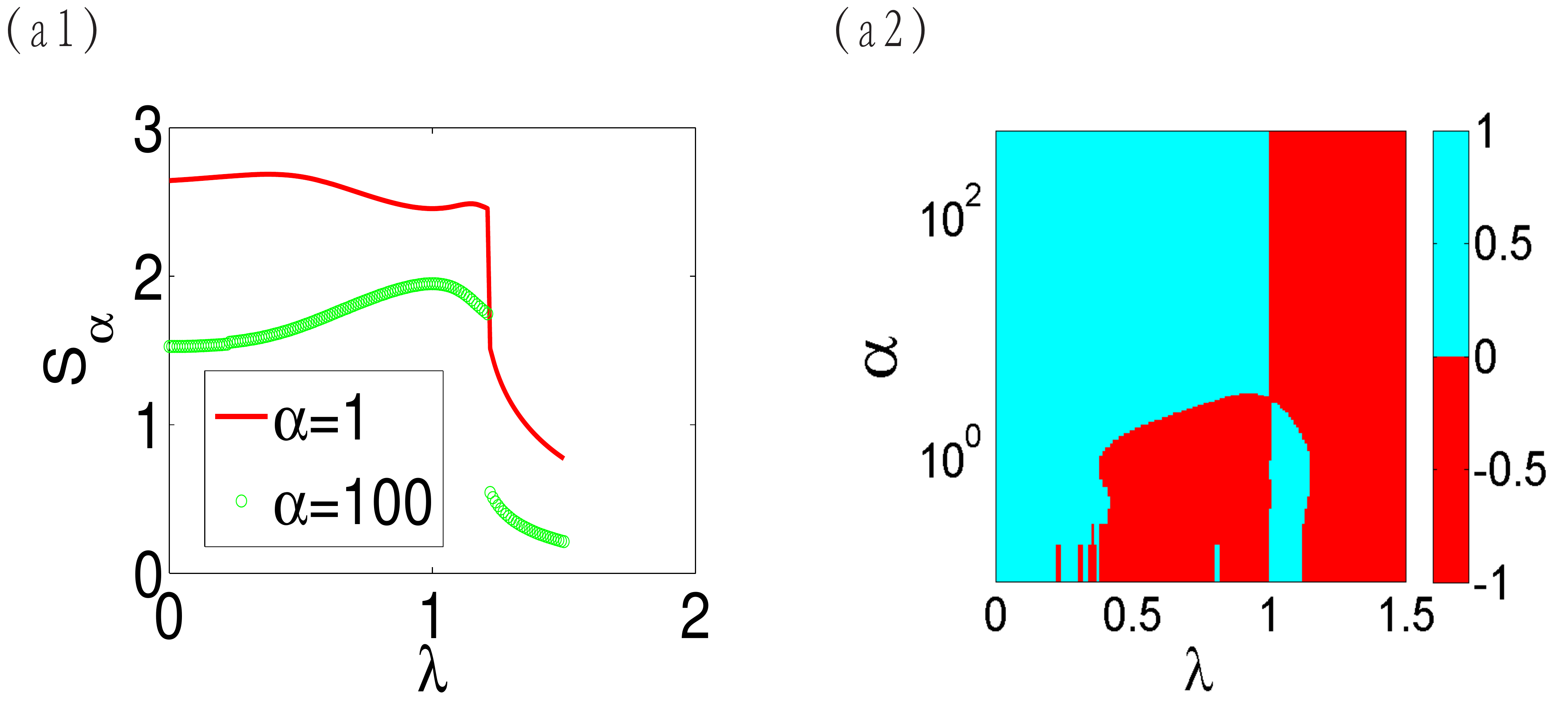}
	\end{minipage}\hfill
	\begin{minipage}[c]{0.42\textwidth}
		\caption{\label{fig_sweep2} 
			Sweep (2) through the $\lambda-D$ model: $D=0$, $\lambda\in \{0,1.5\}$ (see Fig.\ref{sweeps}  for a schematic phase diagram). The sign distribution of the derivative of the R\'{e}nyi entropies $\partial_\lambda S_\alpha$  for partitions $A|B|A$,  $A=48$ and $B=4$ (upper panels) and $A=45$ and $B=10$ (lower panels); both with $N=100$ are presented in $(a2)$. The features of differential local convertibility are characterized by the slopes of the R\'enyi entropies and correspond to specific features of the entanglement spectrum as explained in Fig.\ref{convert_cluster}. The $S_\alpha$'s are presented in $(a1)$ for  $\alpha=100,0.2$ increasing from low to top.  All such quantities are calculated for  the ground  state in $S_z^{tot}=1$ sector}
		\label{convert_large_lambda-D_2}
	\end{minipage}
\end{figure}

In Fig.\ref{sweeps}, we display the schematic phase diagram of the $\lambda-D$ model.
We sweep through  the  phase diagram  in the following two ways: ){\it 1)} fix $\lambda=1$ and change $D$; the Haldane phase is approximately located in the range $-0.4\lesssim D\lesssim 0.8$.  ({\it 2)} Fix $D=0$, varying on $\lambda$; the Haldane phase is located in the range   $0\lesssim \lambda\lesssim 1.1$ (see Fig.\ref{sweeps}). 
We analyzed the four states separately adding the perturbation to the Hamiltonian
$\sim (S_1^z\pm S_N^z)$ with a small coupling constant to resolve the ground state degeneracy.

We find that the N\'eel, ferromagnetic, and the large $D$ phases are locally convertible (see Fig. \ref{conver_lambdaD_symm}: $(a1)$,  $(a2)$).
Consistently with \cite{sym_protect_entang_spectr},
all of the Haldane ground states  are characterized by doubly degenerate entanglement spectrum for the symmetric $A|B$ partitions with $A=B$, for both sweep ways (Fig. \ref{conver_lambdaD_symm}:  $(a3)$ and $(a4)$) (see \cite{vondelft} for an understanding of doubly degenerate entanglement spectrum).  Such a property is not recovered in the cases of asymmetric $A|B$ and $A|B|A$ partitions: in these cases the entanglement spectrum is not found doubly degenerate, because we broke the link inversion symmetry\cite{sym_protect_entang_spectr} (Fig.\ref{conver_lambdaD_symm}:  $(b3)$,  $(b4)$). See also \cite{dechiara} for an analysis of the entanglement spectrum close to the quantum phase transitions.

We find that the Haldane phase is not locally convertible  (see  Figs.\ref{conver_lambdaD_symm}: $(b1)$, $(b2)$, \ref{fig_sweep1}, and \ref{fig_sweep2}).
We remark that for both ways to partition the system the nonlocal-convertibility phenomenon is found even in the case of sizes of $B$ smaller than the correlation length  $\xi$. {As for the model in eq. \eqref{eq:ham_cluster},   we find that the symmetric bipartition $A=B$ displays local convertibility as a fine-tuned effect, which is broken for generic partitions, see Figs. \ref{fig_sweep1} and \ref{fig_sweep2}.

\section{The Perturbed Toric Code}
\label{convert_pertToric}
\markboth{The perturbed Toric Code}{The perturbed Toric Code}

\begin{table}
	\begin{minipage}[c]{0.4\textwidth}
		\begin{tabular}{c c c c c} 
			\hline\hline                        
			Perturbation $V(\lambda)$ & G.I. & DLC & Exact & $\xi$ \\ 
			\hline                  
			$\sum_se^{-\lambda_s\sum_{i\in s}{\sigma^z_i}}$ & \ding{51} & \ding{51} &  \ding{51} & 0 \\ 
			$\lambda_h\sum_{i\in H}{\sigma^z_i}$ & \ding{51} & \ding{55} &  \ding{51} & $\ne 0$ \\
			$\lambda_z \sum_{i}{\sigma^z_i}$ & \ding{51} &  \ding{55} & \ding{55} &$\ne 0$\\
			$\lambda_z\sum_{i}{\sigma^z_i}+\lambda_x\sum_{j}{\sigma^x_j}$ & \ding{55} & \ding{55}   & \ding{55} & $\ne 0$ \\ [1ex]      
			\hline 
		\end{tabular}
	\end{minipage}\hfill
	\begin{minipage}[c]{0.52\textwidth}
		\caption{List of the various perturbation used in the topological Toric Code and some of their properties.
		DLC, i.e no splitting of the R\'enyi entropies, only occurs if the perturbation is fine tuned in order to keep the system with $\xi=0$.
		The left column shows the type of perturbation studied. The first column details whether the considered model is Gauge Invariant. The second column  indicates whether dLC occurs. For certain perturbations the ground state of the system is accessible exactly (third column). The last column provides the information on $\xi$.}
		\label{table}
	\end{minipage}
\end{table}	
	
We now study a set of spin-$1/2$ localized at the edges of a $2D$ square lattice with periodic boundary conditions in presence of a perturbation $V$: \begin{equation}
		\mathcal{H} = - \sum_s  \prod_{i \in s} \sigma^x_i - \sum_p \prod_{i \in p} \sigma^z_i + V(\lambda)
		\label{model}
\end{equation}
where $s$ and $p$ label the vertices and plaquettes of the lattice, respectively, while   $\sigma^x_i$ and $\sigma^z_i$ are Pauli operators of the spin living at  the edge $i$. 
For $V(\lambda)=0$  the Hamiltonian above is the celebrated toric code, a paradigmatic model for topological order\cite{Kitaev}. For the analysis below, we remark that in this case, the ground 
state of this model features  $\xi=0$. We consider different  $V(\lambda)$  (see Table \ref{table}) where $\lambda$ stands for $\{ \lambda_1,\dots,\lambda_N\}$ which are the  parameters controlling the perturbation.  The perturbation in (\ref{model}) is such that the correlation length is increasing with $\lambda_i$ until it divergences at a critical point $\lambda_c$. For a discussion of this criticality, see \cite{critical}.
	
\begin{figure}[b!]
	\begin{minipage}[c]{0.5\textwidth}
		\includegraphics[width=\textwidth,clip=true]{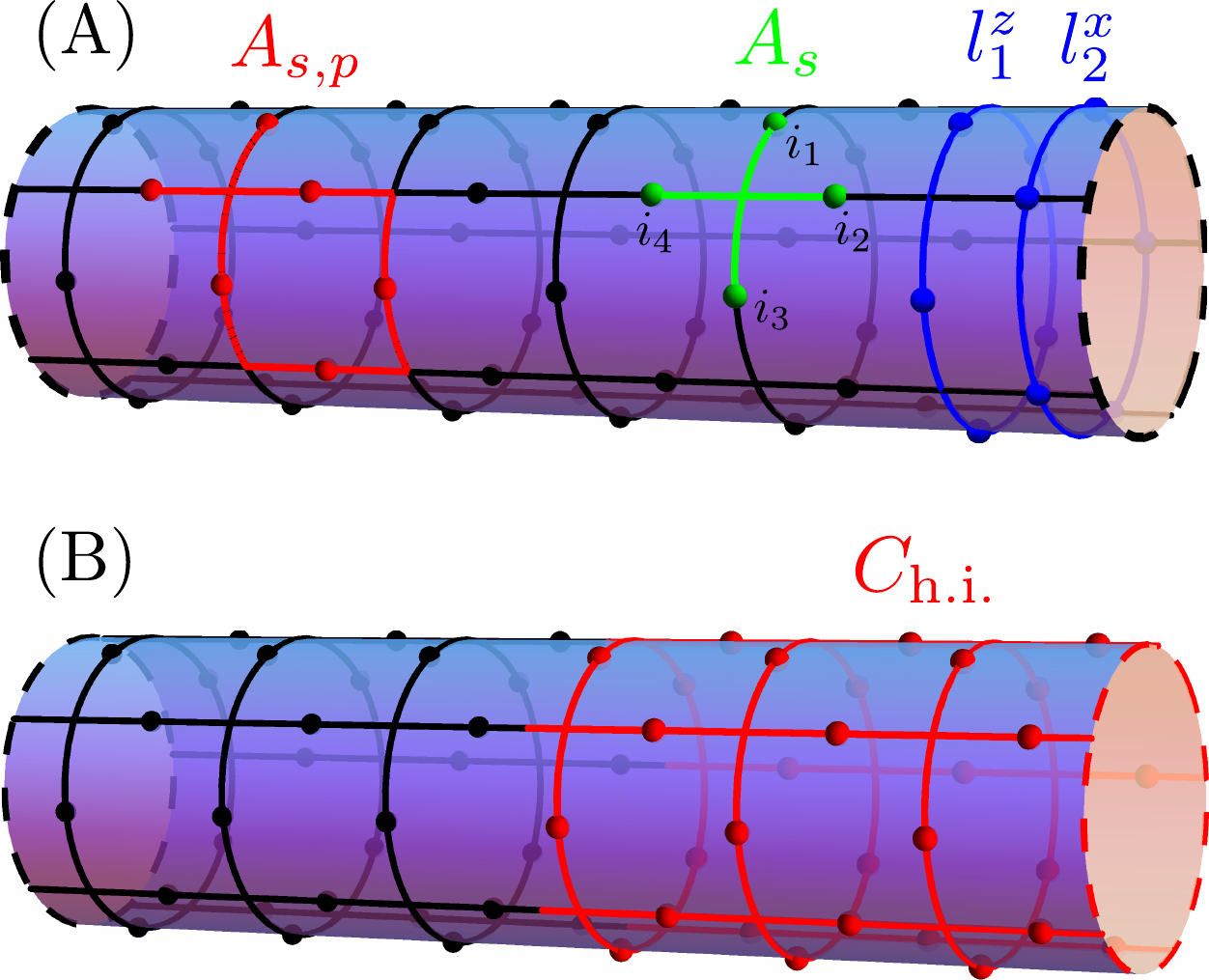}
	\end{minipage}\hfill
	\begin{minipage}[c]{0.47\textwidth}
		\caption{Cylinder of infinite length and width $L_y = 5$ used in 2D DMRG calculation. (a) Subsystems on which R\'enyi entropies are  calculated: $A_s$ - one star and $A_{s,p}$ - composition of star and plaquette. Loops $l^z_1$ and $l^x_2$ used to distinguished between topological sectors are also depicted. (b) Subsystem $C_\textrm{h.i.}$ that contains half of the infinite cylinder.}
		\label {cylinder} 
	\end{minipage}
\end{figure}

For $\lambda<\lambda_c$, these systems are topologically ordered, while for $\lambda>\lambda_c$, they are trivial paramagnets. In both phases, there is no local order parameter. This model belongs to a class of so-called quantum double models that correspond to those phases whose low-energy theory is a lattice gauge theory \cite{Kitaev}.
We demonstrate that one can distinguish the topological from the paramagnetic phases of (\ref{model}) using dLC, even when small subsystems $A$ are considered.

\begin{figure}[t!]
	\begin{minipage}[c]{0.55\textwidth}
	\includegraphics[width=\textwidth,clip=true]{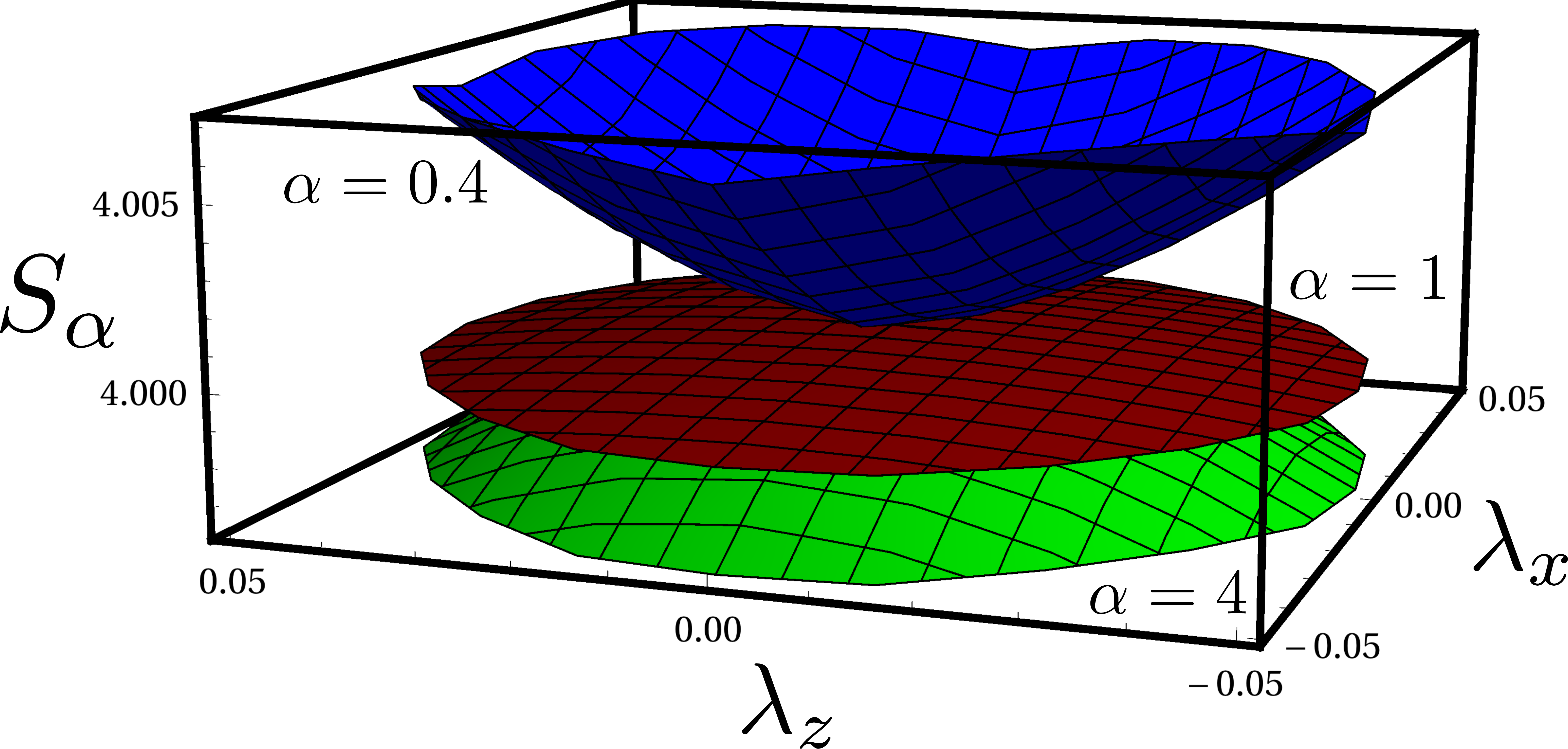}
\end{minipage}\hfill
\begin{minipage}[c]{0.42\textwidth}
	\caption{The splitting phenomenon. The figure displays the splitting with opposite slopes between the small and large $\alpha$ R\'enyi entropies. We see the splitting occurring around $\alpha \simeq 0.6$. The R\'enyi entropies are calculated for the partition $A_{s,p}$ of Fig. \ref{cylinder}a  for the ground state  of $\mathcal H= \mathcal H_{TC} + V_{xz}$}
	\label{3dplot} 
\end{minipage}
\end{figure}

For each $V(\lambda)$ we compute the ground state wavefunction $|\psi(\lambda)\rangle$ and its reduced density matrix $\rho_A (\lambda)$. For some $V(\lambda)$ we can apply  exact analytical approach; for  the  generic perturbation $V_{xz}(\lambda)= \sum_i (\lambda_z {\sigma^z_i}+\lambda_x {\sigma^x_i})$ we  resort to numerical analysis.  
		
The numerical method employed here is an infinite DMRG algorithm \cite{DMRG} in two dimensions. The method provides Matrix Product State (MPS) representation of a complete set of ground states on a cylinder of infinite length and finite width $L_y$ (Fig.\ref{cylinder}) for a given Hamiltonian that realizes topological order. As argued in \cite{Cin12a}, each ground state has a well-defined flux threading through the cylinder. The flux is measured by (in general) dressed Wilson loop operators that enclose the cylinder in the vertical direction.
		
In the case of fixed-point toric code (Eq. \eqref{model} with $V=0$), these loops are given by $l^z_1$ and $l^x_2$ (Fig. \ref{cylinder}a). Four topological sectors are then distinguished by $\langle l^z_1 \rangle, \langle l^x_2 \rangle = \pm 1$. Once the perturbation is present, Wilson loops may change, but as long as the perturbation is small, $\langle l^z_1 \rangle$ and $ \langle l^x_2 \rangle$ can still be used to identify topological sectors because $\langle l^z_1 \rangle, \langle l^x_2 \rangle \simeq \pm 1$.
		
Simulations are carried out with cylinders of width up to $L_y = 5$ for $\sqrt{\lambda_x^2 + \lambda_z^2} \leq 0.05 $ {and $0 \leq \lambda < 0.7$ as shown in Figs. \ref{3dplot} and \ref{p2} respectively}. {In the topological phase,  the} outcome of each simulation is four quasi-degenerate ground states, from which the one with $\langle l^z_1 \rangle, \langle l^x_2 \rangle \simeq + 1$ is chosen for further investigation. {This is done to ensure that finite size effects have the least possible impact on results. In the limit $L_y \rightarrow \infty$ all four ground states become locally indistinguishable.  The results are converged in bond dimension of MPS which acts as a refinement parameter. A reduced density matrix of a half-infinite cylinder $C_\textrm{h.i.}$ (Fig. \ref{cylinder}b) is calculated throughout the simulation. The bond dimension is increased until convergence of its spectrum is reached. 
			
\begin{figure}[t!]
	\begin{minipage}[c]{0.5\textwidth}
		\includegraphics[width=\textwidth,clip=true]{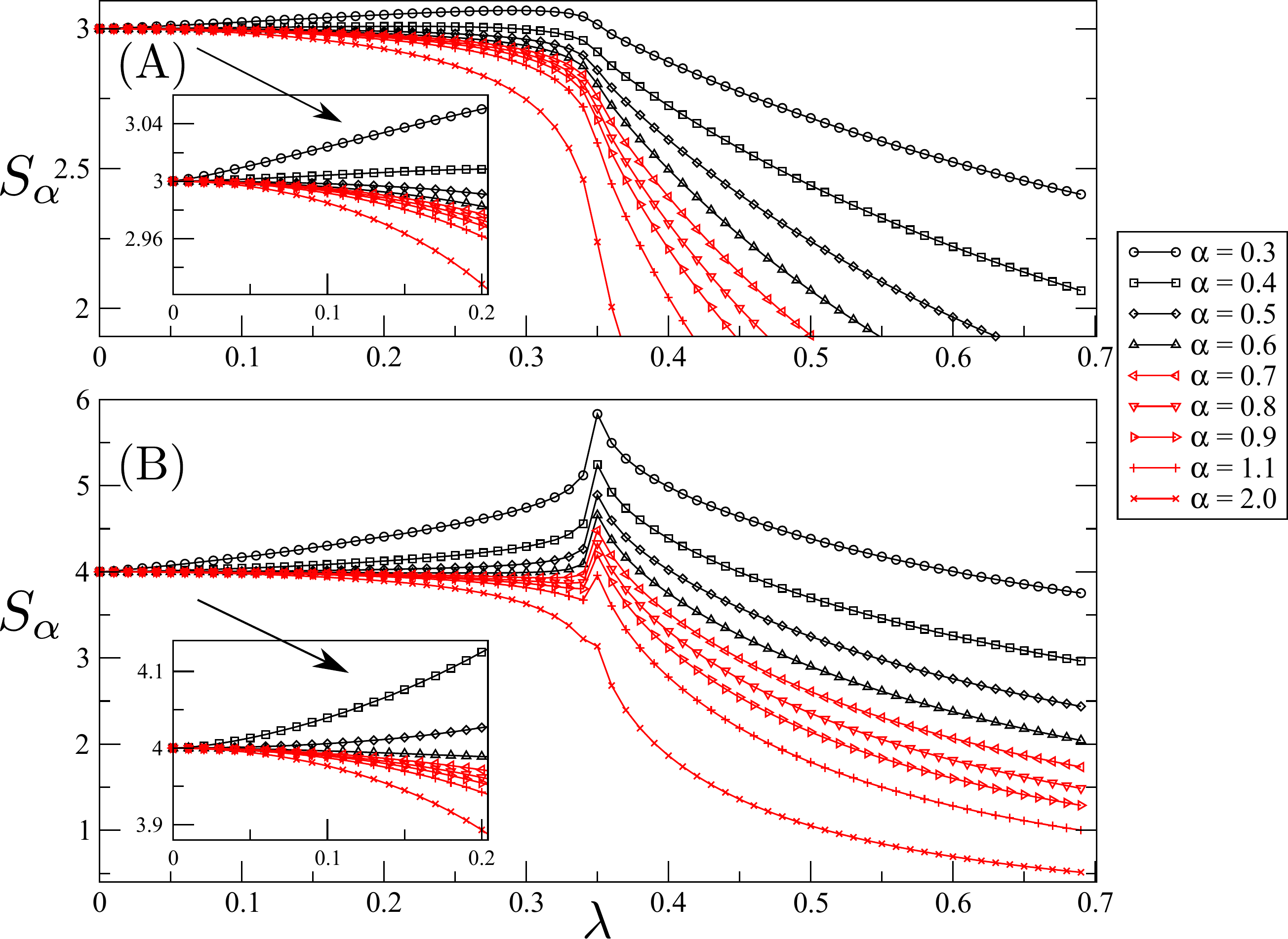}
	\end{minipage}\hfill
	\begin{minipage}[c]{0.47\textwidth}
		\caption{R\'enyi entropies as a function of $\lambda$ for the ground state of $\mathcal H= \mathcal H_{TC} + V_{xz}$ with {$\lambda_x=\lambda$ and $\lambda_z = \lambda/2$}.  Here, $L_y=5$. The reduced system $A$ consists of $A_s$ and $C_\mathrm{h.i.}$ in panels A and B, respectively. As $\lambda$ increases, the correlation length increases. The Schmidt rank $R$ and the low $\alpha<\alpha_0$-R\'enyi entropies increase as well. The value of $\alpha_0$ is $0.4$ and $0.6$ in panels A and B, respectively. Nevertheless, the total entanglement $S_1$ and all the higher R\'enyi entropies are decreasing with $\xi$. Notice the spike in panel B marking the quantum phase transition to the paramagnetic phase at $\lambda_c \sim 0.35$}
		\label{p2} 
	\end{minipage}
\end{figure}
				
In Fig.\ref{3dplot}, we can see the behavior of the $S_\alpha$ R\'enyi entropies as we span the parameter space $\lambda_x, \lambda_z$ for the perturbation $V_{xz}$. {We see clearly that in the topologically ordered phase  a splitting of $S_\alpha$'s occurs:   $\partial_\lambda S_\alpha \lessgtr 0$ at a given value of $\alpha=\alpha_0$;  we found $\alpha_0\simeq 0.6$} (see the caption of Fig. \ref{3dplot}). We will henceforth refer to this phenomenon as  $\alpha$ splitting. In  the paramagnetic phase all the R\'enyi entropies are monotones with $\lambda$. 
This behavior is generically independent of the size and shape of the subsystem $A$, as long as $A$ contains some bulk\cite{Santra2014}. 
Below we explain the phenomenon. The topologically ordered phase we consider is characterized by the presence of a state (at $\lambda=0$) with  $\xi=0$ and a flat entanglement spectrum (and an area law)\cite{renyi}. The flat entanglement spectrum implies that small perturbations result in decreasing $S_\alpha$ for $\alpha > \alpha_0$ being $\alpha_0 < 1$, because the distribution becomes less flat {in the most represented eigenvalues in the entanglement spectrum}. In contrast, $S_0$ must increase with $\xi$ as an effect of the perturbation (new degrees of freedom are involved in the entanglement spectrum). 
So the $\alpha$ splitting results from the insertion of a finite $\xi$ in the state evolving from a state with a flat spectrum and zero $\xi$.
We also observe that such property is shared with the so-called $G$-states that include all the topologically ordered quantum double models and states like the cluster states\cite{hiz3}, and therefore our findings apply to this class of models as well\cite{renyi} (see \cite{kalis} for a discussion of the cluster phase diagram).  Here we remark that the splitting effectively distinguishes a class of quantum spin liquids (states with finite correlation length and no local order parameter), which are notoriously very difficult to detect, since one cannot measure correlation functions of all the possible local observables. To further distinguish non-topologically ordered quantum spin liquids like the cluster states from topologically ordered states, we need to measure the degeneracy of the ground state, since the former have a unique ground state while topological states possess a degeneracy protected by topology. 
				
Moreover, notice that the splitting occurs no matter how we perturb in the plane $\lambda_x, \lambda_z$, and it is, therefore, a robust property of the phase.  Note again that in the paramagnetic phase all the $\partial_\lambda S_\alpha$ have the same sign and no splitting ever occurs, which is easily understood from the presence (at very large $\lambda$) of a completely factorized state, see Fig.\ref{p2}.
				
We remark that the splitting phenomenon effectively distinguishes the topologically ordered state from a topologically trivial ordered state (like a ferromagnet). As discussed above, the latter states have typically $S_\alpha$ increasing with $\xi$ and no splitting occurs. 
Summarizing, we can distinguish between the topological phase and the paramagnet of (\ref{model}); furthermore, we can distinguish between the topological phase and a symmetry-breaking phase.
				
To corroborate our findings, we resort to exact analysis for suitable perturbations $V(\lambda)$'s. 
We consider two cases {\it i)} $V_h = \lambda_z \sum_{i\in h} \sigma_i^z$, corresponding to placing the  external field $\propto \sigma^z$ only along the horizontal links of the lattice; and {\it ii)} $V(\lambda)= \sum_{s}e^{-\lambda\sum_{i \in s}{\sigma}^z_i}$ leading to the Castelnovo--Chamon model\cite{chamon}. Since these perturbations commute with the plaquette operators of Eq. (\ref{model}), the ground state of these models can be written as the superposition of  loop states $\ket{g}$ with amplitudes $\alpha(g)$. A loop state $\ket{g}$ is obtained from the completely polarized state in the $z$ direction, by flipping down all the spins intersected by a loop drawn on the lattice. The corresponding loop  operators $g$ form a group $G$ called the gauge group of these theories.

In case {\it i)}  the  star operators $\prod_{i \in s } \sigma^x_i$ interact only along the rows of the lattice. The model maps onto the product of arrays of Ising chains by the duality $A_s\to \tau^z_{\mu}$, $\sigma^z_i\to \tau^x_{\mu}\tau^x_{\mu+1}$:  $\mathcal H_{TC} + V_h \mapsto \mathcal{H}_{ff}=\bigoplus_{i=1}^L(-\lambda\sum_{\mu}\tau^x_{\mu}\tau^x_{\mu+1}-\sum_{\mu}\tau^z_{\mu})$ \cite{Dusuel,Halasz}. The relevant correlators in the variables $\sigma$ can be obtained through the correlators in the dual variables $\tau$ that can be accessed exactly\cite{mcoy}. In the following, we sketch a proof that the splitting phenomenon does occur in this model (see \cite{Santra2014} for additional details). 
We consider the star  $A_s=\{ i_1, i_2, i_3, i_4\}$ as subsystem A (see Fig. \ref{cylinder});   $\rho_{A_s}$ is {block diagonal with $2 \times 2$ blocks labeled by $\ket{i_1i_2 i_3 i_4}$ and $A_s\ket{i_1i_2 i_3 i_4}$}. It results that $\rho_A$ has maximum rank unless $\alpha(g) =\alpha(g_1)\alpha(g_2)$, implying there is a zero eigenvalue in each block. 
In the dual picture this  is equivalent to require $\langle \tau_i \tau_j\rangle= \langle \tau_i \rangle \langle \tau_j\rangle$. Such condition holds at 		$\lambda=0$ only, and therefore  $R$ increases at  $\lambda\neq0$}. 
The factorization of the amplitudes also proves that both $\alpha=1,2$-R\'enyi entropies decrease at small $\lambda$ \cite{Santra2014}. 
			
The case  ({\it ii)}  is important to test the argument of the interplay between splitting and increasing of the correlation length. This argument implies that
a perturbation for which $\xi(\lambda)=const$ does not lead to a splitting in the R\'enyi entropies. The model of Castelnovo--Chamon  features exactly this since spin--spin correlation functions $\langle \sigma^x_i \sigma^x_j\rangle$ are vanishing for every value of $\lambda$.
The exact ground state is made of loops with amplitudes $\alpha(g)= e^{-\lambda/2\sum_{i\in s}\sigma^z_i(g)}$, where $\sigma^z_i(g) =  \bra{g} \sigma^z_i\ket{g}$. 
The  topological phase is ($\lambda<\lambda_c\approx 0.44$). 
A lengthy calculation leads to  
$
S_{\alpha}(\rho_A)=(1-\alpha)^{-1}\log {Z^{-\alpha}(\lambda)}\sum_{g\in G}e^{-\lambda L_g}w^{\alpha-1}(\lambda,g)
$,
where  $Z= \sum_g e^{-\lambda L_g}$ and $w(\lambda,g):=\sum_{h\in G_A,k\in G_B}e^{-\lambda L_{hgk}}$, and $L_m$ is the length of the loop  $m$ of the gauge group $G$; here $G_A$ and $G_B$ are the gauge groups of the subsystems $A$ and $B$, respectively.
The analysis   of  small and large $\lambda$ expansions reveals that $\partial_{\lambda}S_{\alpha}(\lambda)\le 0$\cite{Santra2014}. }As a particular case, $S_0$ is constant for every value of  $\lambda$. Accordingly, for this fine-tuned perturbation all R\'enyi entropies decrease and therefore no splitting is observed. This is consistent with the fact that also in this model the amplitudes $ \alpha(g)$ factorize as discussed in ({\it i)}.

\section{The Quantum Ising Chain}
\label{QIsingSec}
\markboth{The quantum Ising chain}{The quantum Ising chain}

To better extract the effect of edge states on local convertibility, it is desirable to have a model with three properties:
(i) it should support edge states, (ii) quasi-particle excitations should be identifiable, and (iii) there should be a mechanism for destroying the edge states and observing the different behavior. 
The one-dimensional transverse field Ising model fulfills these requirements  \cite{lieb1961,mcoy}.
It is defined by the Hamiltonian 
\begin{eqnarray}
H_{\rm I}=-\sum_{j=1}^N \left( t \: \sigma_j^x \sigma_{j+1}^x + h \: \sigma_j^{z} \right) \; ,
\label{HIsing}
\end{eqnarray}
where $t$ is a hopping amplitude (which we can set to $t=1$) and $h$ is the control parameter for the external magnetic field. A quantum phase transition for $h=t=1$ happens in the thermodynamic limit of $N \to \infty$.  This QPT's signatures have been recently observed experimentally \cite{coldea2010}.

The Hilbert space of (\ref{HIsing}) can be described in terms of eigenstates of the string operator $\mu_N^x = \prod_{j=1}^N \sigma_j^z$, which generates the $\mathbb{Z}_2$ symmetry of (\ref{HIsing}). For $h>1$ the system is paramagnetic with $\langle \sigma^x \rangle =0$. For $h<1$, the spectrum of the Ising model becomes doubly degenerate. A ground state that is also an eigenstate of $\mu_N^x$ has a vanishing order parameter $\langle \sigma_x \rangle =0$. This ground state is known as the ``thermal ground state''. This is the state employed in the  2-Sat problem and adiabatic quantum computation protocols for finite $N$ \cite{adiabatic}. In the thermodynamic limit ($N \to \infty$),  $\sigma^x$ can acquire a nonzero expectation value. The symmetry will be broken spontaneously and the ground state will be given by the (anti)symmetric combination of the two eigenstates of $\mu_N^x$. For $h<1$  we consider both the ferromagnetic ground state (MSBGS) with nonvanishing order parameter $\langle \sigma^x\rangle$ and the thermal one enjoying the same  $\mathbb{Z}_2$ symmetry as the Hamiltonian.

The quantum Ising model (\ref{HIsing}) can be mapped exactly, although nonlocally, to  a system of free spinless fermions $\{c_j,c^\dagger_j\}$, see  \cite{lieb1961}. We remark that this mapping preserves the entanglement between $A$ and $B$ \cite{vidal2003, latorre2004} and generates the  {\it Kitaev chain}. As emphasized in \cite{kitaev2000} this formulation highlights the presence of Majorana edge states as emergent degrees of freedom. Majorana fermions are the elusive particles (coinciding with their own anti-particles), proposed by E. Majorana. Many research groups are trying to find and manipulate them \cite{hasan2010, mourik2012}.
Each Dirac  fermion of the chain can be used to define two Majorana fermions:
\be
  f_j^{(1)} \equiv \left[ \prod_{l<j} \sigma_l^z \right] \sigma_j^x = c_j^\dag + c_j \, , \quad 
  f_j^{(2)} \equiv \left[ \prod_{l<j} \sigma_l^z \right] \sigma_j^y =\ii \left( c_j^\dag - c_j \right) \; 
\ee

\begin{figure}[t]
	\begin{minipage}[c]{0.55\textwidth}
		\includegraphics[width=\textwidth]{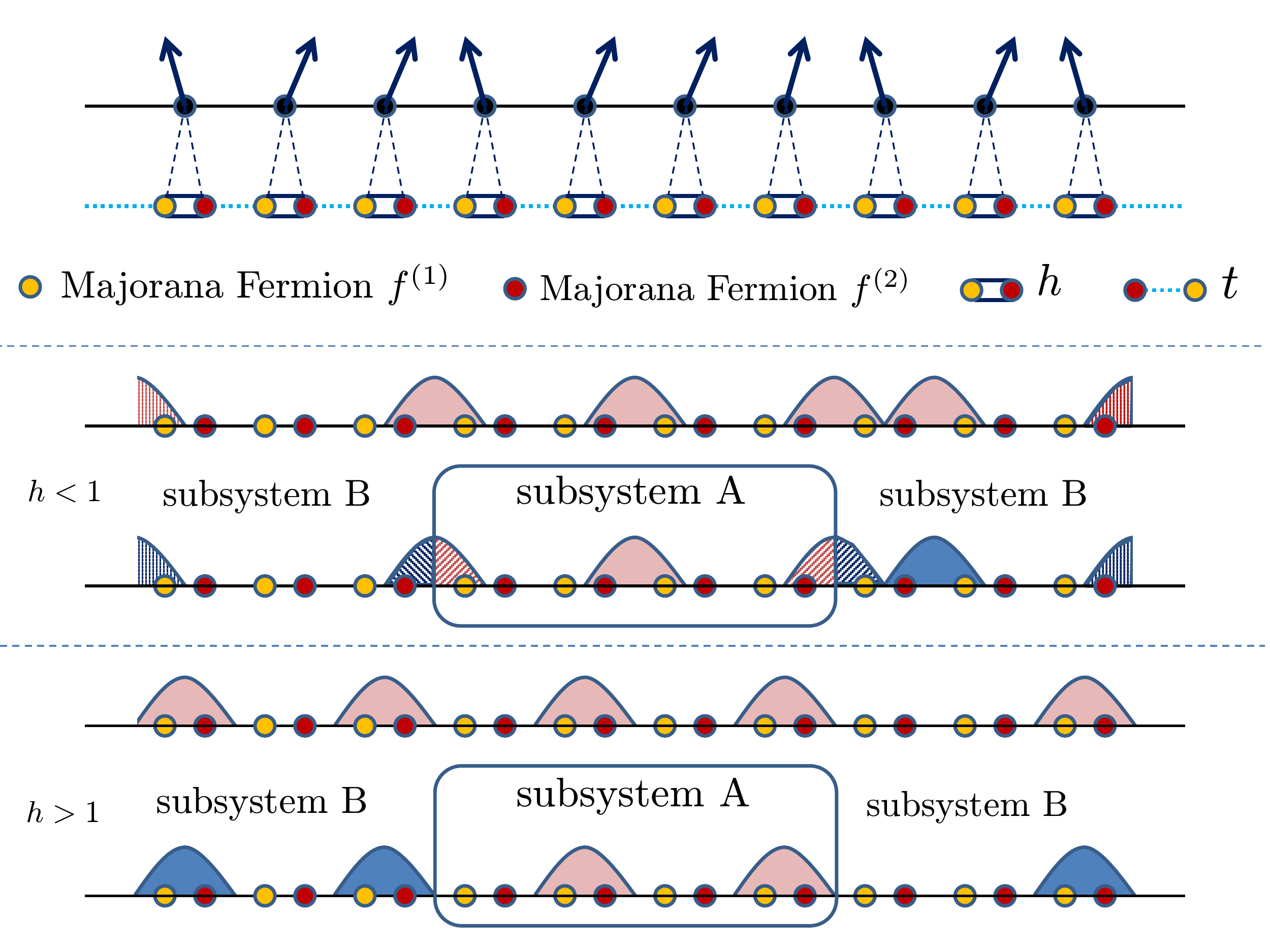}
	\end{minipage}\hfill
	\begin{minipage}[c]{0.42\textwidth}
		\caption{Top: the Ising chain is mapped into a system of Majorana fermions by doubling the lattice sites. Middle and bottom: a schematic cartoon of the  quasi-particle excitations in the two phases of the model and the effect of bipartitioning the system; for small $h$,  edge states form at the opposite  boundaries of the subsystem A. The property of local convertibility depends on the correlations between such edge states}
		\label{Majorana}
	\end{minipage}
\end{figure}

We represent this mapping pictorially in Fig. \ref{Majorana}. In the paramagnetic phase ($h>1$), the Hamiltonian pairs predominantly Majoranas on the same site $j$ (this correlation is drawn as a double line in the picture). In the ferromagnetic phase ($h<1$), the dashed line connecting different sites is dominant. In Kitaev's approach, the double degeneracy of this phase emerges as the first and last Majoranas are left unpaired and can be combined into a complex fermion (the occupancy/vacancy of this fermion costs no energy). We will see that the same picture applies when the system is divided into two partitions: in the ferromagnetic phase this operation cuts the dominant link and leaves unpaired Majorana edge states on each side of the cut.

This is a key many-body feature that renders phases supporting boundary states more  ``quantum'' than other systems and hence more powerful when employed as simulators. Since any subsystem develops its edge states, in these phases q-bits of information are stored nonlocally between the sites and we will see that this is mirrored by the nontrivial entanglement behavior, yielding nonlocal convertibility. Such phenomenology, which can hardly be implemented in a classical setting, must be a fundamental ingredient of a machine aimed at simulating a generic quantum system and this is the reason for which nonlocal convertibility is a strong indicator of a higher computational power.

\subsection{The R\'enyi Entropies}

An advantage of working with a quadratic theory such as the Ising chain is that many-body states can be constructed exactly out of individual quasi-particle excitations. The latter can be found as the linear combination of the fermionic operators $\{ c_j , c_j^\dag \}$ which diagonalizes the Hamiltonian. Doing so, we define a new set of operators $\{ \tilde{c}_j , \tilde{c}_j^\dag \}$ so that  the ground state $|0\rangle$ is  annihilated by all $\tilde{c}_j$. On top of it, one can excite quasi-particles by progressively applying all possible combinations of $\tilde{c}_j^\dag$, giving a total of $2^N$ states in the Hilbert space. 

To calculate the entanglement between the subregions $A$ and $B$, we use the Schmidt decomposition of the ground state
\be
| 0 \rangle = \sum_l \sqrt{\lambda_l} \, | \psi_l^{(A)} \rangle \, | \psi_l^{(B)} \rangle \; ,
\ee
where $|\psi_l^{(A,B)} \rangle$ span the Hilbert space of block $A$ and $B$ respectively \cite{NielsenChuang}. We are after the eigenvalues $\lambda_l$, which can be found, for instance, as
\be
\lambda_l = \langle 0 | \psi_l^{(A)} \rangle \langle \psi_l^{(A)} | 0 \rangle \; ,
\label{lambdaproj}
\ee
where a tracing over the $B$ degrees of freedom is implicitly assumed.
Similarly to what is done for the whole system, the states $|\psi_l^{(A)}\rangle$ can be constructed in terms of individual excitations. However, these are different from those of the whole chain, as they are completely contained inside the block. If $A$ consists of $L$ consecutive sites, these block excitations $\{ d_j , d_j^\dag \}$ are the linear combinations of the $c$-operators within the block, which diagonalize the correlation matrix constructed out of all their two-point correlation functions, as shown below. Each state $|\psi_l^{(A)}\rangle$ of this $2^L$-dimensional Hilbert space can thus be characterized by the occupation number $0$ or $1$ of each block excitation. Moreover, the eigenvalues $\nu_j$ of the aforementioned correlation matrix provide us with the expectation values
\be
\langle 0| d_j d_j^\dagger | 0 \rangle = {1 + \nu_j \over 2} \; , \qquad
\langle 0| d_j^\dagger d_j | 0 \rangle = {1 - \nu_j \over 2} \; ,
\label{Bmodes}
\ee
all other correlations being zero. Note that $\nu_j \simeq 1$ indicates that $d_j$ annihilates the vacuum $|0\rangle$. It follows that certain quasi-particle excitations of the Hamiltonian are completely contained within the block, since $d_j | 0\rangle = 0 $ implies that $d_j$ is just a superposition of $\tilde{c}_j$'s. Since $d_j$ is defined just within the block, it follows that these $\tilde{c}_j$'s are also contained in the block. Conversely, smaller values of $\nu_j$ are related to excitations lying only partially within a subregion.
In turn, $d_j d_j^\dag$ acts on the ground state as a projection operator which selects the component with $0$ occupation number for the $l$th block excitation, while $d_j^\dag d_j$ projects it on an occupied $l$th excitation. Hence, (\ref{lambdaproj}) can be written as the expectation value of a string of operators of this type. Using (\ref{Bmodes}) as the building blocks of these correlators, we have
\be
\{ \lambda_l \} = \prod_{j=1}^L \left( {1 \pm \nu_j \over 2} \right) \, ,
\label{lambdadef}
\ee
with all the possible combinations of plus/minus signs, corresponding to the occupation/unoccupation of the different block excitations. 

Finally, the R\'enyi entropies read  \cite{vidal2003, latorre2004}
\be
S_\alpha (\rho_A) = {1 \over 1- \alpha} \sum_{j=1}^L \log \left[
\left(  {1 + \nu_j \over 2} \right)^\alpha
+ \left(  {1 - \nu_j \over 2} \right)^\alpha \right] \; .
\label{Salpha}
\ee

\subsection{The Correlation Matrix}

As we just discussed, the R\'enyi entropies are accessed through the ``eigenvalues'' of the reduced density matrix of a block of $L$ consecutive spins for the thermal ground state \cite{vidal2003, latorre2004}. Such ``eigenvalues'' can be obtained from the diagonalization of the $2L \times 2L$ correlation matrix: $ \langle f_k^{(a)} f_j^{(b)} \rangle =\delta_{j,k} \delta_{a,b} + i \left( {\cal B}_L \right)_{(j,k)}^{(a,b)}$, with
\be
{\cal B}_L \equiv \left( \begin{array}{cccc}
	\Pi_0 & \Pi_1 & \ldots & \Pi_{L-1} \cr
	\Pi_{-1} & \Pi_0 && \vdots \cr
	\vdots & & \ddots & \vdots \cr
	\Pi_{1-L} & \ldots & \ldots & \Pi_0 \cr
\end{array} \right),
\label{Bdef}
\ee
where $j,k$ specifies the entry $\Pi_{j-k} \equiv \left( \begin{array}{cc} 0 & g_{j-k} \cr - g_{k-j} & 0 \cr \end{array} \right)$, which is itself a $2 \times 2$ matrix whose $a,b$ entries are defined as
\be
g_j \equiv {1 \over 2 \pi} \int_0^{2 \pi} { \cos \theta -h  + \ii \sin \theta \over
	\sqrt{ (\cos \theta - h )^2 +  \sin^2 \theta} } \: \eu^{\ii j \theta} \: \de \theta \: .
\label{gjdef}
\ee

The antisymmetric matrix ${\cal B}$ can be brought into a block-diagonal form by a $SO(2L)$ rotation, with each block of the form $$\tilde{\Pi}_j = \nu_j \left( \begin{array}{cc} 0 & 1 \cr - 1 & 0 \cr \end{array} \right)$$ This rotation defines a new set of Majorana fermions $\tilde{f}_j^{(a)}$ with only pair-wise correlations. This rotated operator basis  can be used to introduce the new set of complex operators: $d_j = \left( \tilde{f}_j^{(1)} + \ii \tilde{f}_j^{(2)} \right)/ 2$ (and their Hermitian  conjugated). The matrix (\ref{Bdef}) contains all the information to completely solve the model. By taking $L =N$, i.e. extending the correlation matrix to the whole system, the $d$-modes coincide with the $\tilde{c}$-operators, one would obtain from the diagonalization of the Hamiltonian.

For $L=2$, the two eigenvalues of the correlation matrix are easily found to be
\be
\nu_\pm = \sqrt{ \left({g_1 - g_{-1} \over 2} \right)^2 + g_0^2}  \pm {g_1 + g_{-1} \over 2} \; ,
\label{nupm}
\ee
which allows for a complete analytical study of the entanglement entropy and its derivative (see Fig. \ref{2spinAn}).

\subsection{The $\mathbb{Z}_2$ Symmetric Ground State}

Thus, in the Ising chain, the $2^L$ states within a block of $L$ consecutive sites can be constructed in terms of individual quasi-particle excitations, which can be either occupied or empty. 
These excitations are in general delocalized, with a typical size set by the correlation length. However, a  $\mathbb{Z}_2$ symmetric state possesses one special excitation, with support lying at the opposite edges of the block and formed by two Majorana edge states \cite{kitaev2000}. When the block is extended to the whole system ($L=N$), the block excitations coincide with the system's excitations, including the boundary states.

The entanglement between two subsystems $A$ and $B$ can be extracted from the $2 L$ eigenvalues $\pm \ii \nu_j$ of the correlation matrix Eq.(\ref{Bdef}) incorporating the correlations of the excitations within the spin block. Here $L$ is the number of lattice sites in $A$. The eigenvalues of the reduced density matrix can then be constructed out of the $\nu_j$'s, using (\ref{lambdadef}) in the {\it method section}. The $\nu$'s can be interpreted as sort of {\it occupation numbers}, since they capture the overlaps between each block quasi-particle excitation and the ground state, according to (\ref{Bmodes}): $\nu_j =0$ means that this block excitation is half-filled and half-empty in the ground state, while $\nu_j = 1$ indicates that the excitation is either completely occupied or not present at all.

\begin{figure}[t]
	\begin{center}
	\includegraphics[width=.4\textwidth]{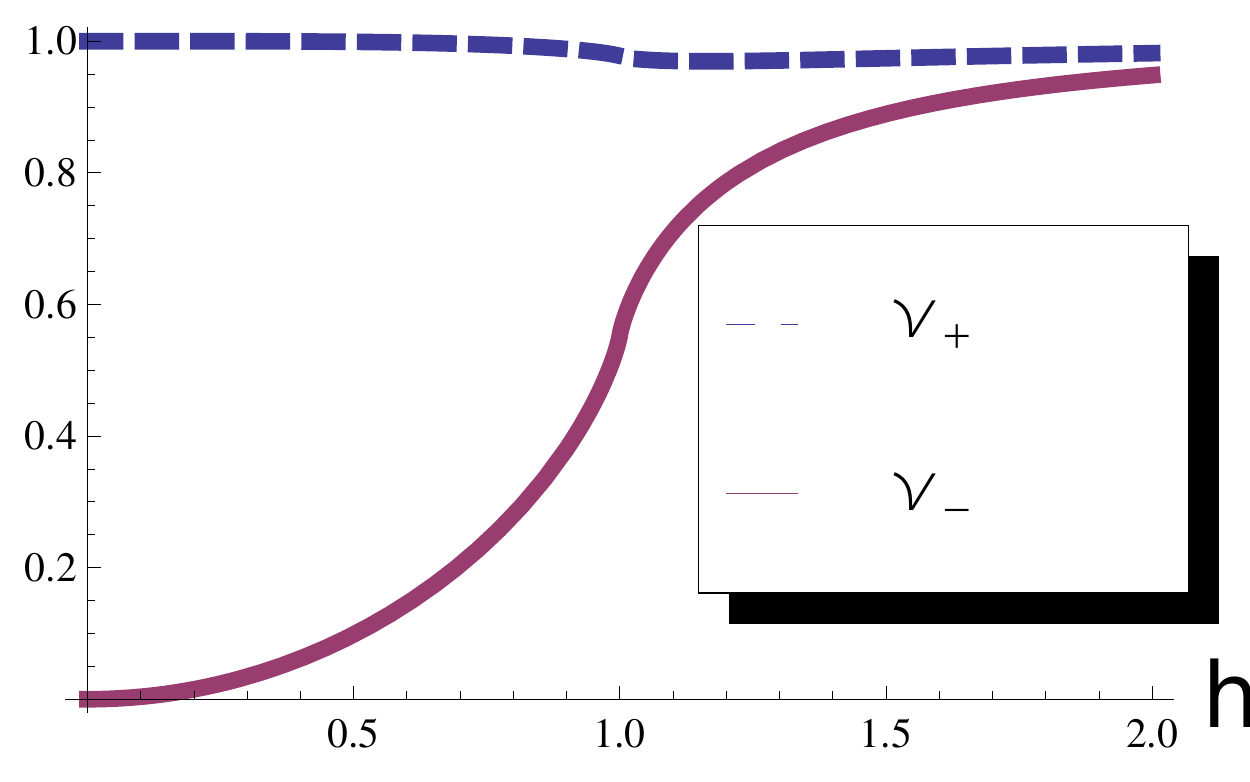} \hskip1cm
	\includegraphics[width=.4\columnwidth]{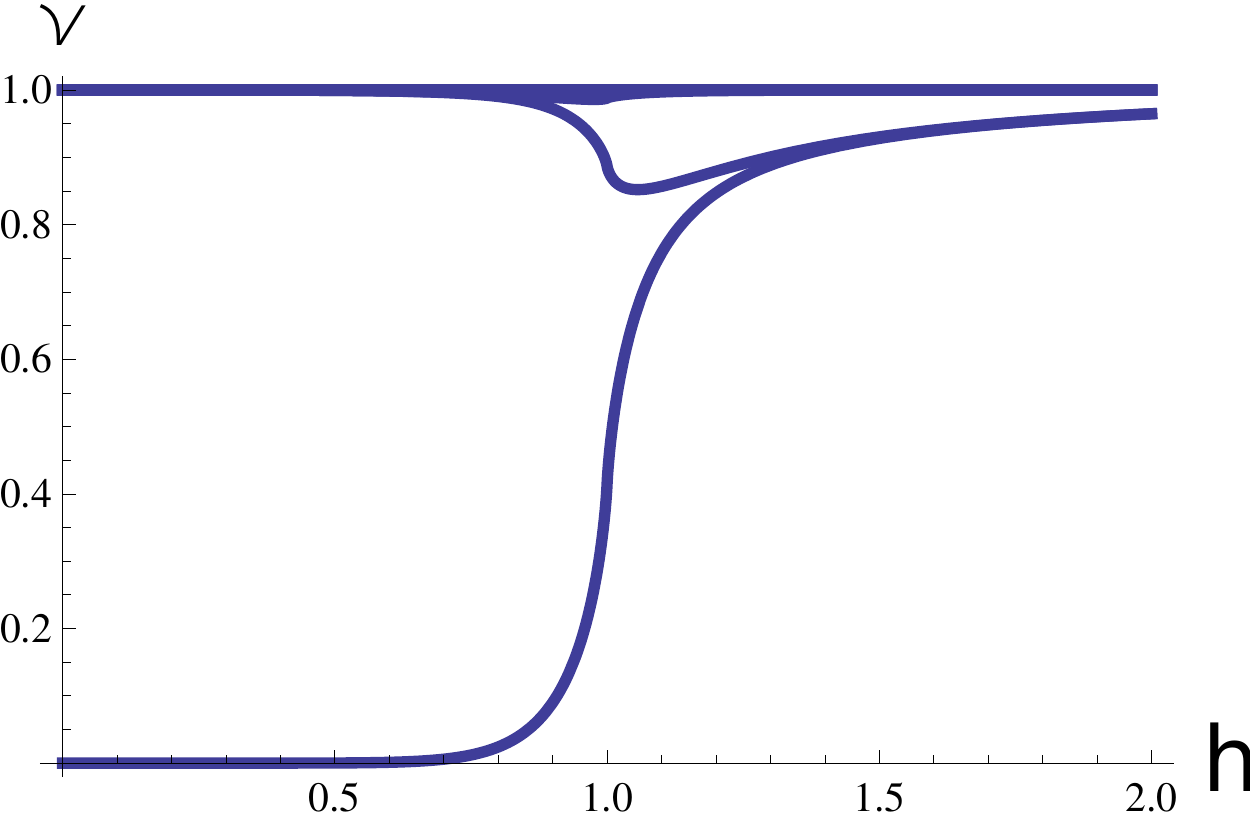}
	\end{center}
	\caption{Plot of the occupation number $\nu_j$ obtained from the correlation matrix (\ref{Bdef}) as a function of $h$ for $L=2$ (left) and $L=10$ (right). For $L=2$, the explicit form of the eigenvalues $\nu_\pm$ is given in (\ref{nupm}). Notice that only one of the $\nu$'s shows a nontrivial behavior: it corresponds to the boundary state, which is only partially contained in the subregion}
	\label{fig:correigen}
\end{figure}

In Fig. \ref{fig:correigen} we plot these eigenvalues $\nu_j$ as a function of the magnetic field for $L=2$ and $L=10$. Notice that in both cases only one block excitation has a nontrivial behavior, while the other eigenvalues stay approximately constant around unity in all phases.  Significant deviations happen only close to the QPT (as the correlation length diverges). As discussed, the modes with $\nu_j \simeq 1$ define {\it bulk states}.
In contrast, the nontrivial eigenvalue is close to zero for $h \simeq 0$ and rises rapidly toward $1$ crossing the QPT at $h=1$: in the ferromagnetic phase, it corresponds to a block excitation which is neither occupied nor empty. By cutting the chain into two subregions, we severed the dominant inter-site correlation and hence generated two unpaired Majorana edge states (see Fig. \ref{Majorana}). We noticed, however, that as $h$ increases, the occupation number of this edge excitation increases, indicating edge state recombination.

Having discussed the behavior of the eigenvalues $\nu_j$'s and the role of the boundary states, it is straightforward to analyze the R\'enyi entropy and address the issue of differential local convertibility.
It is interesting to concentrate on the two extreme limits: $L=2$ and $L \to \infty$.

\begin{figure}[t]
	\begin{center}
	\includegraphics[width=.4\textwidth]{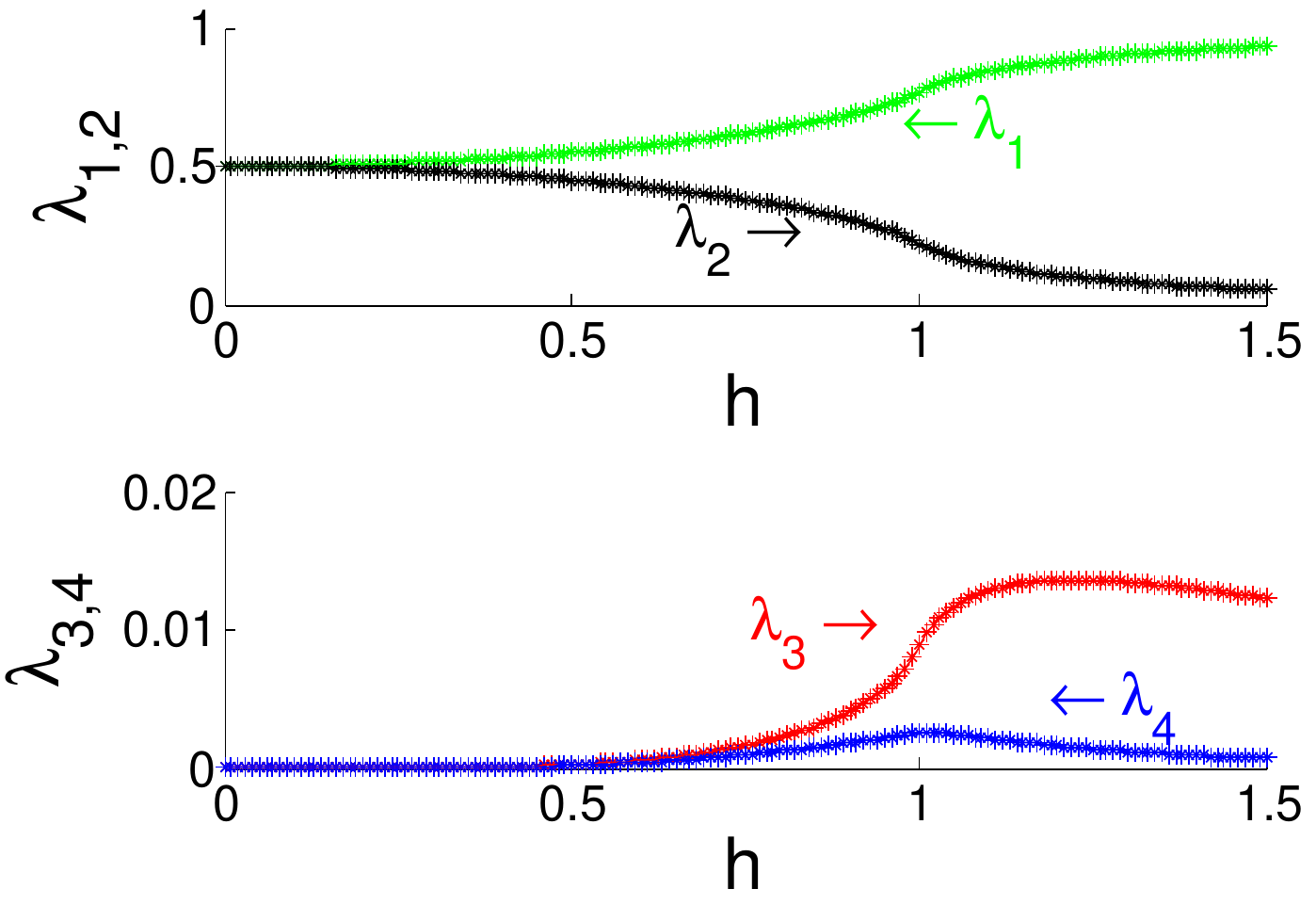} \hskip1cm
	\includegraphics[width=.4\columnwidth]{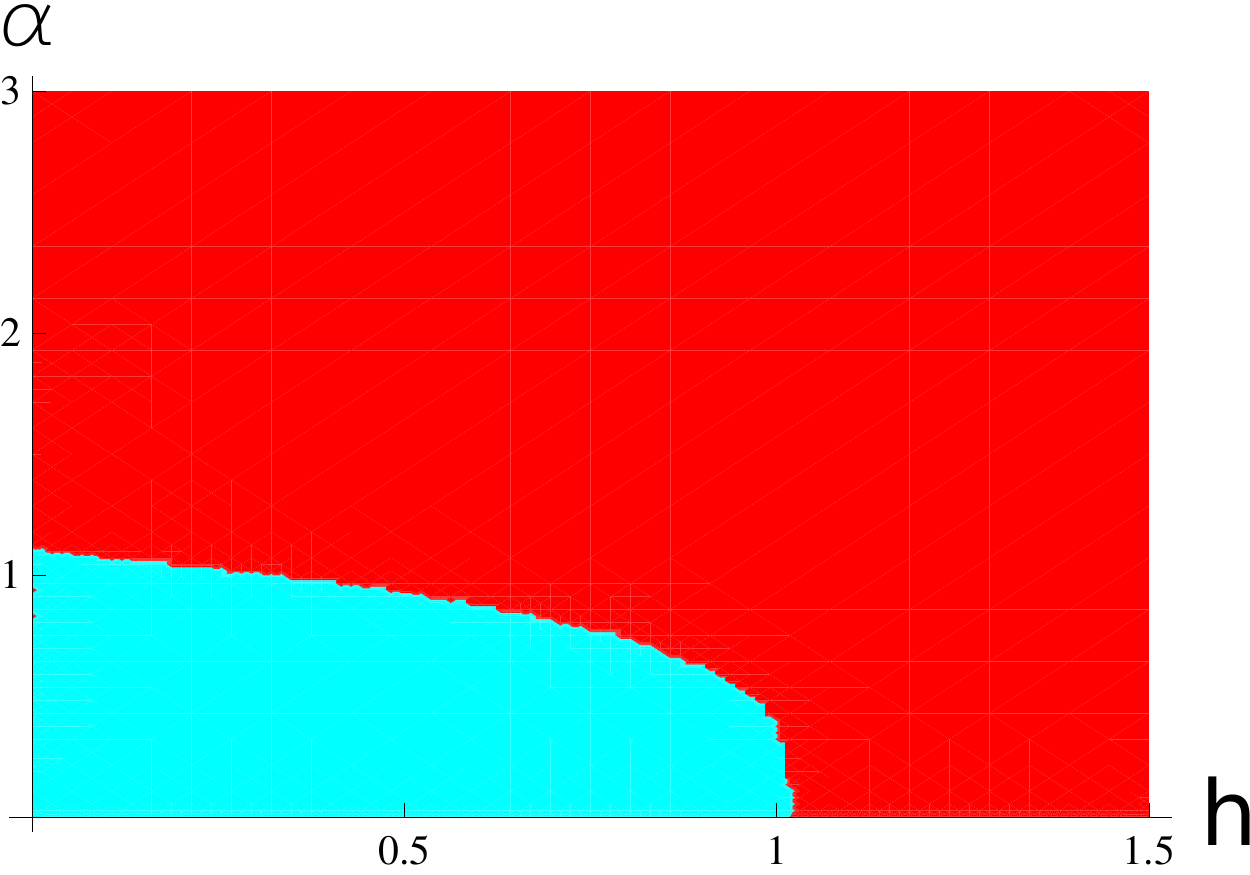}
	\end{center}
	\caption{Left: plot of the four eigenvalues of $\rho_A$ for $L=2$, as a function of $h$. The solid lines are the analytical results, while the crosses show the numerical ones with N=200 (notice the different scales in the vertical axis between the top and bottom panels.). Right: contour plot of the sign of the derivative by $h$ of the R\'enyi entropy for different values of $h$ and $\alpha$}
	\label{2spinAn}
\end{figure}

The two occupation numbers $\nu_\pm$ for $L=2$ are shown in the left panel of Fig. \ref{fig:correigen} and the resulting four eigenvalues of the reduced density matrix, according to (\ref{lambdadef}), are plotted in the left panel of Fig. \ref{2spinAn}. While in locally convertible phases the largest eigenvalue(s) decrease approaching the QPT, indicating an increase of the entanglement; here, we see that the edge state recombination results in a growing larger eigenvalue. The right panel of  Fig. \ref{2spinAn} presents the sign of the entanglement entropy derivative, to be considered with dLC.
We see that in the paramagnetic phase the R\'enyi entropy always decreases. Instead, in the doubly degenerate phase, the entropy derivative vanishes at some value of $\alpha$ and changes sign, indicating that local (differential) convertibility is lost in this phase (as already observed numerically for small $N$ and larger $L$ in \cite{Cui-Gu}).
It is important to notice here that these results imply that classical local gates operating on two sites project out the half-occupied state and hence lose the edge state entanglement.

\begin{figure}[t]
	\begin{center}
	\includegraphics[width=.4\textwidth]{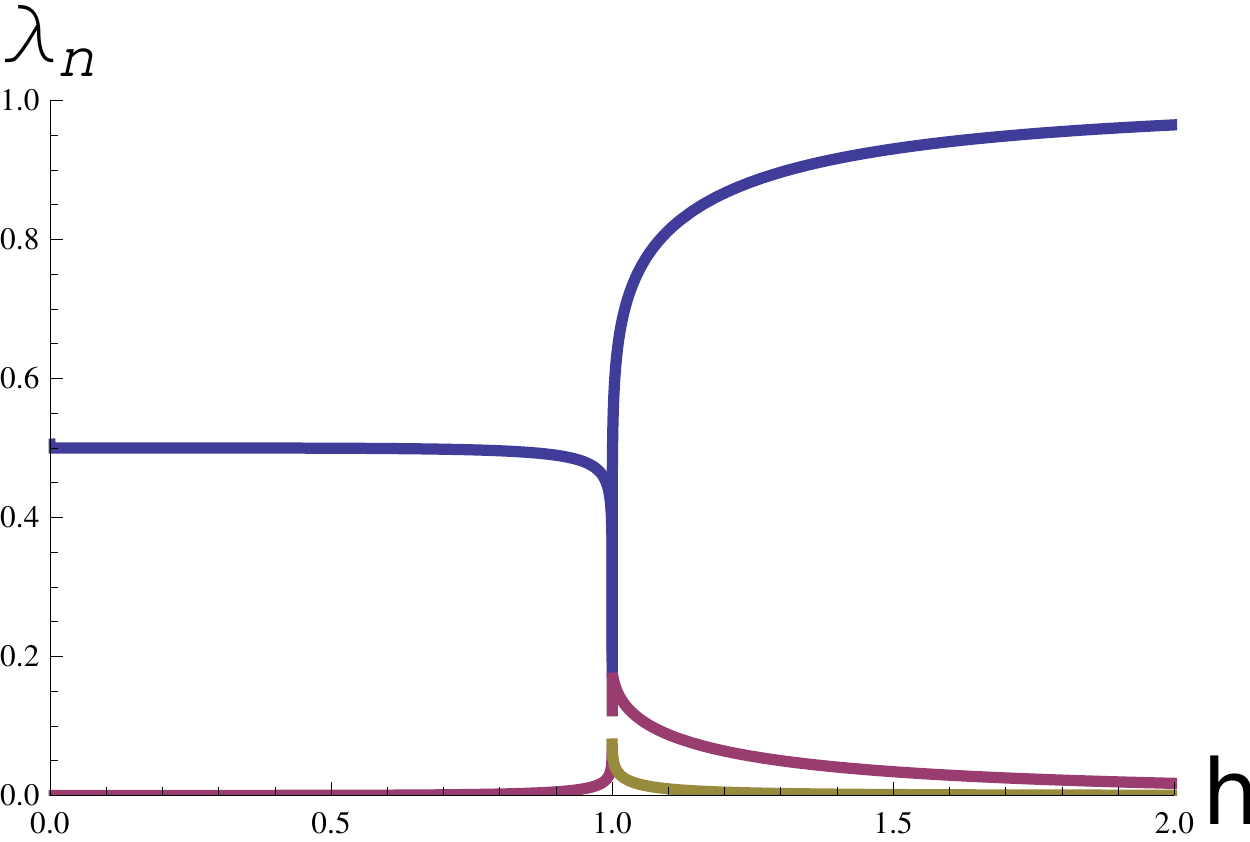} \hskip1cm
	\includegraphics[width=.4\columnwidth]{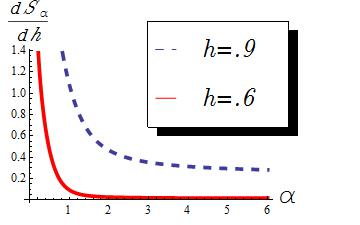}
	\end{center}
	\caption{Left: plot of the first few eigenvalues of $\rho_A$, for an infinite size block, as a function of $h$. The eigenvalues' multiplicities are not shown (for instance, the highest eigenvalue is doubly degenerate for $h<1$ and unique for $h>1$, see \cite{franchini10}). Right: plot of the derivative of the R\'enyi entropy with respect to the magnetic field $h$, as a function of $\alpha$, for two different values of $h$ in the ferromagnetic region}
	\label{InfBlockAn}
\end{figure}

For the $L \to \infty$ limit, we can take advantage of the results of \cite{its05, franchini08, franchini10}, where the full spectrum (eigenvalues and multiplicities) of the reduced density matrix and the R\'enyi entropies were calculated analytically. Figure \ref{InfBlockAn} shows a plot of the first few eigenvalues of $\rho_A$ and a plot of the entropy derivative as a function of $\alpha$  for $h=0.6$ and $h=0.9$. We see that the largest eigenvalue (doubly degenerate in the ferromagnetic phase) decreases monotonously toward the QPT, while smaller eigenvalues are allowed to grow, yielding a monotonous increase of all the R\'enyi entropies. It is thus clear that local convertibility is restored in the infinite $L$ limit.

We checked these results numerically for systems up to $N=200$ and with different partitions. We considered different block sizes and move the blocks within the chain. The qualitative picture does not change significantly as one varies $(A|B)$, but the location of the curve where the entropy derivative vanishes moves in the $(h,\alpha)$ space. It tends toward the phase transition line $h=1$ as the block sizes grow bigger, confirming our expectation on the role of the boundary excitations. Namely, we see that as long as the edge states from different boundaries do not overlap, their occupation number eigenvalue stays constant and vanishes.  It starts increasing only once the correlation length grows comparable to one of the block sizes, {\it indicating the recombination of the edge states and a decrease in the entanglement contribution from the edge states}.

\subsection{Symmetry Broken Ground State}

\begin{figure}[t]
	\begin{center}
	\includegraphics[width=.4\textwidth]{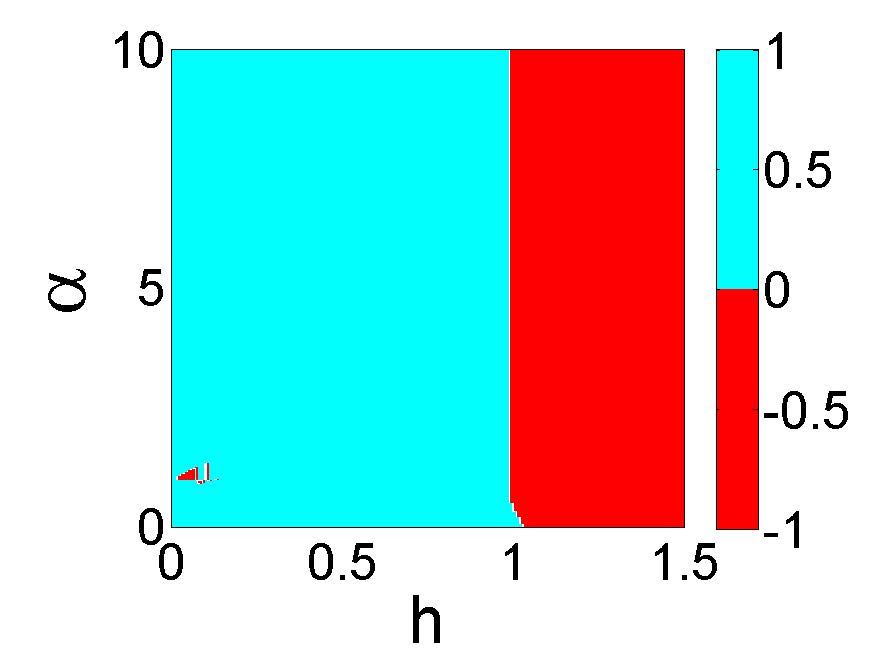} \hskip1cm
	\includegraphics[width=.4\columnwidth]{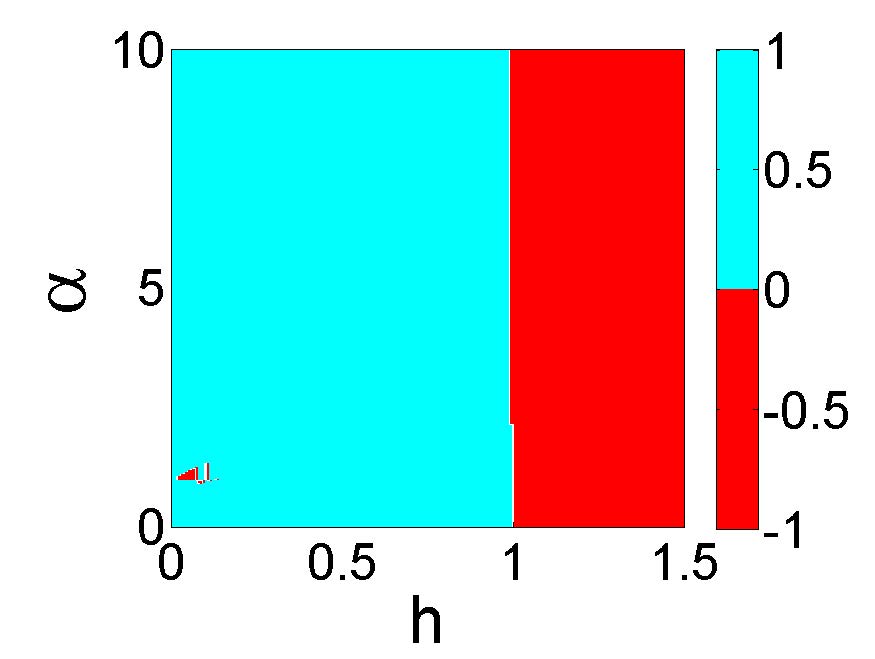}
	\end{center}
	\caption{Numeric results of differential local convertibility for the ferromagnetic (symmetry broken).
	On the left, a partition $200=2|198$ and on the right $200=50|100|50$}
	\label{symmetrybroken}
\end{figure}

To further confirm our interpretation on the role of boundary modes, in the ordered phase $h<1$, we also considered the ferromagnetic ground state for which $\langle \sigma^x \rangle \neq 0$. Since this state does not support well-defined Majorana edge states, we expect a restoration of local convertibility.
We calculate the R\'enyi entropy of this symmetry broken ground state numerically. Namely, we add a very small perturbation $\epsilon(\sigma_1^x+\sigma_N^x)$ to Hamiltonian (\ref{HIsing}) and apply the variational MPS routine to obtain the ground state. In this work the converge tolerance is $10^{-6}$.
Figure \ref{symmetrybroken} shows the plots of the sign of the entropy derivative for two possible partitions (small and large $A$ block) and validates our expectation that both phases are locally convertible. We considered several partitioning choices and the results are not distinguishable from those in Fig. \ref{symmetrybroken}.

In conclusion, we see that for $h>1$ the disordered ground state is always locally convertible. In the ordered phase, the ferromagnetic ground state, i.e. with broken symmetry, is also locally convertible for any chosen partition. For the thermal ground state, however, the convertibility depends on the interplay between the size of the partitions $(A|B)$  and the correlation length of the system. This phenomenon is a manifestation of edge state recombination.  
These entangled pairs lie on opposite boundaries of the partition (see Fig.\ref{Majorana}), but with finite support intruding in the bulk about the order of the correlation length.
For sufficiently large block sizes, the entanglement between boundary states does not depend on the correlation length and remains constant throughout the phase. 
However,  as this length increases approaching a QPT, the edge states effectively grow closer to one another.  If either of the subregions $A$ and $B$ is sufficiently small, the tails of these states can overlap and we see their occupation number increasing and {\it their entanglement decreasing}, yielding nonlocal convertibility. 

\section{Classical nature of ordered quantum phases and origin of spontaneous symmetry breaking}
\label{classicalsec}
\markboth{Origin of SSB}{Origin of SSB}

To discuss the differences among different ground states of a model with a degenerate lowest energy manifold, we take as an example the ferromagnetic one-dimensional spin-$1/2$ $XY$ in the presence of a transverse field and periodic boundary conditions.
The Hamiltonian of such model reads~\cite{lieb1961,Pfeuty1970,BMD1970,mcoy}
\begin{equation}\label{eq:XYmodelhamiltonian}
	H =- \sum_{i=1}^{N} \left[ \left( \frac{1+\gamma }{2} \right)  \sigma_i^x \sigma_{i+1}^x  + 
	\left( \frac{1-\gamma }{2} \right) \sigma_i^y
	\sigma_{i+1}^y  +  h \sigma_i^z\right]  \;,
\end{equation}
where $\gamma$ is the anisotropy parameter in the $xy$ plane, $h$ is the transverse local field, and the periodic boundary conditions
$\sigma_{N+1}^\mu  \equiv  \sigma_1^\mu$ ensure a perfect invariance under spatial translations.

For this class of models, the phase diagram can be determined exactly in great detail~\cite{lieb1961,BMD1970,Franchini17}. 
In the thermodynamic limit, for any $\gamma \in (0,1]$, a quantum phase transition occurs at the critical value $h_c = 1$ of the transverse field. 
For $h < h_c$ the system is characterized by a bidimensional ground state manifold in which two elements are living in both parity sectors~\cite{VCM2008}. 
As a consequence, in such a manifold, it is possible to define elements showing a ferromagnetic order along the $x$ axis which highlights the fact that they violate the $\mathbb{Z}_2$ parity symmetry. 
Indeed, given the two symmetric ground states, the so-called even $|e\rangle$ and odd $|o\rangle$ states belonging to the two orthogonal subspaces associated with the two possible distinct eigenvalues of the parity operator, any symmetry-breaking linear superposition of the form
\begin{equation}\label{eq:groundstates}
	|g(u,v)\rangle = u |e\rangle + v |o\rangle \;
\end{equation}
is also an admissible ground state, with the complex superposition amplitudes $u$ and $v$ constrained by the normalization condition \mbox{$|u|^2 + |v|^2 = 1$}.
Taking into account that the even and odd ground states are orthogonal, the expectation values of operators that commute with the parity operator are independent of the superposition amplitudes $u$ and $v$. On the other hand, spin operators that do not commute with the parity may have nonvanishing expectation values on such linear combinations and hence break the symmetry of the Hamiltonian (\ref{eq:XYmodelhamiltonian}).

Consider observables $O_S$ that are arbitrary products of spin operators and anti-commute with the parity. Their expectation values in the superposition ground states (\ref{eq:groundstates}) are of the form
\begin{equation}\label{eq:expectationvalue}
	\langle g(u,v)|O_S|g(u,v)\rangle = u v^* \langle o|O_S|e\rangle + v u^* \langle e|O_S|o\rangle \; .
\end{equation}
Both $\langle o|O_S|e\rangle$ and $\langle e|O_S|o\rangle$ are real and independent of $u$ and $v$ and hence the expectation
(\ref{eq:expectationvalue}) is maximum for \mbox{$u = \pm v = 1 / \sqrt{2}$}~\cite{BMD1970}. These are the values of the
superposition amplitudes that realize the maximum breaking of the symmetry and identify the order parameter as well as the MSBGSs.

Besides the quantum critical point, there exists another relevant value of the external magnetic field, that is $h_f=\sqrt{1-\gamma^2}$,
the {\em factorizing field}. Indeed, at this value of $h$, the system admits a twofold degenerate, completely factorized ground
state~\cite{Kurmann1982,Roscilde2005,Giampaolo2008,Giampaolo2009,Giampaolo2010}.

To discuss the entanglement and discord-type correlations of quantum ground states, we consider arbitrary bipartitions $(A|B)$ such that subsystem $A=\{i_1,\ldots,i_L\}$ is any subset made of $L$ spins, and subsystem $B$ is the remainder. Given any global ground state of the total system, the reduced density matrix $\rho_A$ ($\rho_B$) of subsystem $A$ ($B$) can be expressed in general in terms of the $n$-point correlation functions~\cite{Osborne2002}:
\begin{equation}
	\label{eq:defreduce}
	\rho_{A} (u,v)  =   \frac{1}{2^L}        
	\ \sum_{\mu_1,\ldots,\mu_L}       \langle g(u,v) | \sigma_{i_1}^{\mu_1}  \cdots   \sigma_{i_L}^{\mu_L}  
	|g(u,v) \rangle \sigma_{i_1}^{\mu_1}  \cdots   \sigma_{i_L}^{\mu_L} \, ,
\end{equation}
and analogously for $\rho_B$. All expectations in Eq.~(\ref{eq:defreduce}) are associated with spin operators that either commute or anti-commute with the parity along the spin-$z$ direction. Therefore, the reduced density matrix $\rho_A$ can be expressed as the sum of a symmetric part $\rho_A^{(s)}$, i.e., the reduced density matrix obtained from $|e\rangle$ or $|o\rangle$, and a traceless matrix $\rho_A^{(a)}$ that includes all the terms that are nonvanishing only in the presence of a breaking of the symmetry:
\begin{equation}
	\label{eq:defreduce1}
	\rho_{A}(u,v) = \rho_A^{(s)} + (uv^*+vu^*) \rho_A^{(a)} \; .
\end{equation}
Both $\rho_A^{(s)}$ and $\rho_A^{(a)}$ are independent of the superposition amplitudes $u$ and $v$, while the reduced density matrix depends on the choice of the ground state. This implies that the elements of the ground space are not locally equivalent. Explicit evaluation of expectations and correlations in symmetry-breaking ground states in the thermodynamic limit is challenging even when the exact solution for the symmetric elements of the ground space is available.

We will now sketch a method that allows overcoming this difficulty and whose general validity is not in principle restricted to the particular model considered. To obtain $\rho_A^{(s)}$ it is sufficient to transform the spin operators into fermionic ones and then apply Wick's theorem. Such algorithm cannot be applied to spin operators $O_A$, acting on subsystem $A$, that anti-commute with the parity. To treat this case we first introduce the symmetric operator $O_AO_{A+r}$, for which, by applying the previous procedure, we can evaluate $\langle e |O_A O_{A+r}| e\rangle$. Then, the desired expectation $\langle e |O_A| o\rangle$ can be computed by exploiting the property of asymptotic factorization of products of local operators at infinite separation~\cite{BMD1970,S2000,Bratteli2012} that yields $\langle e |O_A| o\rangle = \sqrt{\lim\limits_{r \to \infty} \langle e |O_A O_{A+r}| e\rangle}$,
where the root's sign is fixed by imposing positivity of the density matrix $\rho_{A}(u,v)$. Having obtained the exact reduced density matrix $\rho_{A}(u,v)$ for all possible subsystems $A$ and superposition amplitudes $u$ and $v$, we are equipped to investigate the nature of quantum ground states for their properties of classicality and quantumness.

\subsection{Two-Body Quantum Correlations}
\label{sec:definitionsandnotationsofstellarentandquant}

Among all the different possibilities, in the present section, we focus on the analysis of the behavior of one-way discord-type correlations and entanglement between any two spins for different ground states. 
One-way discord-type correlations are properties of quantum states more general than entanglement. 
Operationally, they are defined in terms of state distinguishability for the so-called {\em classical-quantum} states. 
The latter are quantum states that, besides being separable, i.e. not entangled, remain invariant under the action of at least one nontrivial local unitary operation. 
In geometric terms, a valid measure of quantum correlations must quantify how much a quantum state {\em discords} from classical-quantum states and must be invariant under the action of all local unitary operations. 
A computable and operationally well-defined geometric measure of quantum correlations is then the {\em discord of response}~\cite{Roga2014,Giampaolo2013}. The pairwise discord of response $D_R$ for a two-spin reduced density matrix is defined as
\begin{equation}
	\label{discord}
	D_R(\rho_{ij}^{(r)}(u,v)) \equiv \frac{1}{2} \min_{U_i} d_{x} \left(\rho_{ij}^{(r)}(u,v),\tilde{\rho}_{ij}^{(r)}(u,v) \right)^2 \, ,
\end{equation}
where $\rho_{ij}^{(r)}(u,v)$ is the state of two spins $i$ and $j$ at a distance $r$, obtained by taking the partial trace of
the ground state $|g(u,v)\rangle$ with respect to all other spins in the system, $\tilde{\rho}_{ij}^{(r)}(u,v) \equiv  U_i\rho_{ij}^{(r)}(u,v) U_i^\dagger$ is the two-spin state transformed under the action of a local unitary operation $U_i$ acting on spin $i$, and $d_{x}$ is any well-behaved, contractive distance (e.g. Bures, trace, Hellinger) of $\rho_{ij}^{(r)}$ from the set of locally unitarily perturbed states, realized by the least-perturbing operation in the set. The trivial case of the identity is excluded by considering only unitary operations with {\em harmonic} spectrum, i.e., the fully nondegenerate spectrum on the unit circle with equispaced eigenvalues.

For pure states, the discord of response reduces to an entanglement monotone, whose convex-roof extension to mixed states is the so-called \textit{entanglement of response}~\cite{Giampaolo2007, Monras2011, Gharibian2012}. Therefore, the entanglement and the discord of response quantify different aspects of bipartite quantum correlations via two different uses of local unitary operations. The discord of response arises by applying local unitaries directly to the generally mixed state, while the entanglement of response stems from the application of local unitaries to pure states. Under their common origin, it is thus possible to perform a direct comparison between these two quantities.

In terms of the trace distance, which will be relevant in the following, the two-qubit entanglement of response is simply given by the squared concurrence~\cite{Wootters1998, Roga2014}, whereas the two-qubit discord of response relates nicely to the trace distance-based geometric discord~\cite{Nakano2013}, whose closed formula is known only for a particular class of two-qubit states~\cite{Ciccarello2014}, although it can be computed for a more general class of two-qubit states through a very efficient numerical optimization.

\paragraph{Symmetry-Preserving Ground States}\label{sec:symmetricgroundstate}

\begin{figure}[t]
	\includegraphics[width=8cm]{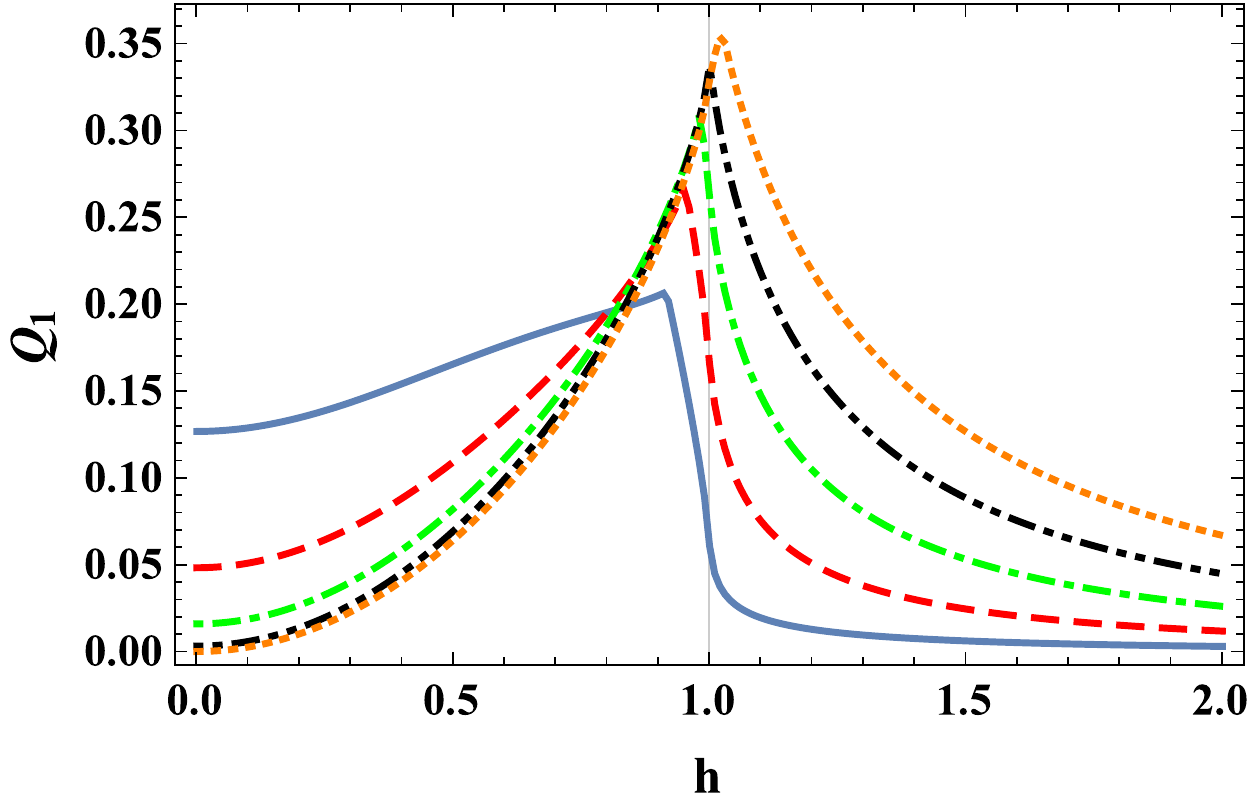}
	\includegraphics[width=8cm]{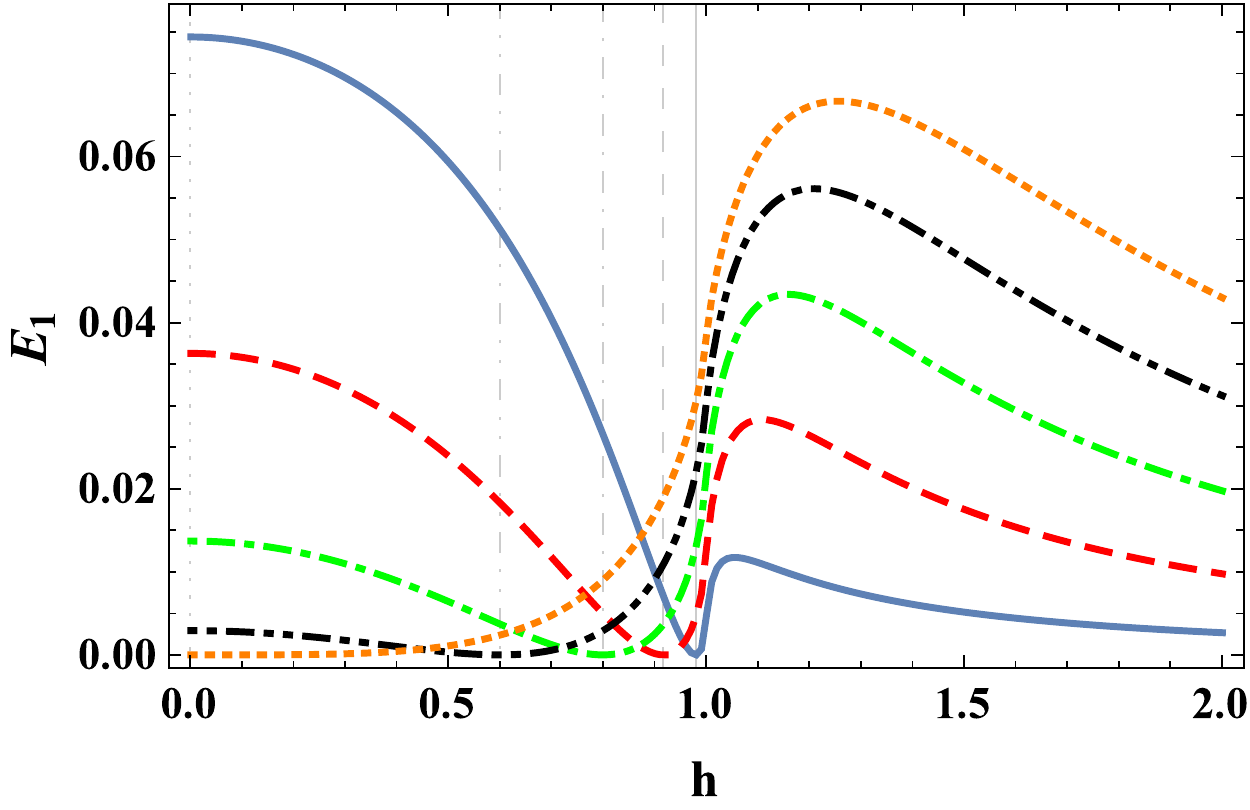}
	\caption{Nearest-neighbor trace distance-based discord of response (left panel) and nearest-neighbor trace distance-based entanglement of response (right panel) for symmetry-preserving ground states, in the thermodynamic limit, as functions of the external field $h$, and for different values of the anisotropy $\gamma$. Solid blue curve: $\gamma=0.2$; dashed red curve: $\gamma=0.4$; dot-dashed green curve: $\gamma=0.6$; and double-dot-dashed black curve: $\gamma=0.8$; dotted orange curve: $\gamma=1$. In the lower panel, to each of these curves, there corresponds a vertical line denoting the associated factorizing field $h_f$. In the upper panel, the solid vertical line denotes the critical field $h_c = 1$}
	\label{fig:symnearestquantandent}
\end{figure}

We first compare the two-body entanglement of response and the two-body discord of response in symmetry-preserving ground states.
For two neighboring spins, these two quantities are plotted in Fig.~\ref{fig:symnearestquantandent} as functions of the external field $h$ and for different values of the anisotropy $\gamma$. For any intermediate value of $\gamma$, the nearest-neighbor entanglement of response $E_1$ exhibits the following behavior. If $h<h_f$, $E_1$ decreases until it vanishes at the factorizing field $h=h_f$. Otherwise, if $h>h_f$, $E_1$ first increases until it reaches a maximum at some value of $h$ higher than the critical point $h_c=1$, and then it decreases again until it vanishes asymptotically for very large values of $h$ in the paramagnetic phase (saturation). Overall, $E_1$ features two maxima at $h=0$ and $h>h_c$ and two minima at $h=h_f$ (factorization) and $h\rightarrow\infty$ (saturation). For the Ising model ($\gamma=1$) the point $h=0$ corresponds instead to a minimum,
since it coincides with the factorizing field $h_f=\sqrt{1-\gamma^2}$.

On the other hand, regardless of the value of $\gamma$, the nearest-neighbor discord of response $Q_1$ always features a single maximum. Depending on the value of $\gamma$ such maximum can be either in the ordered phase $h<h_c$ or in the disordered (paramagnetic) phase $h>h_c$, moving toward higher values of $h$ with increasing $\gamma$. Remarkably, $Q_1$ never vanishes at the factorizing field, except in the extreme case of $\gamma=1$. Indeed, at the factorizing field $h=h_f$, and for any $\gamma\neq 0,1$, the symmetry-preserving ground state is not completely factorized but rather
is a coherent superposition with equal amplitudes of the two completely factorized MSBGSs. Consequently, while the two-body entanglement of response must vanish by the convex roof extension, the two-body discord of response remains always finite.

\begin{figure}[t]
	\includegraphics[width=7.65cm]{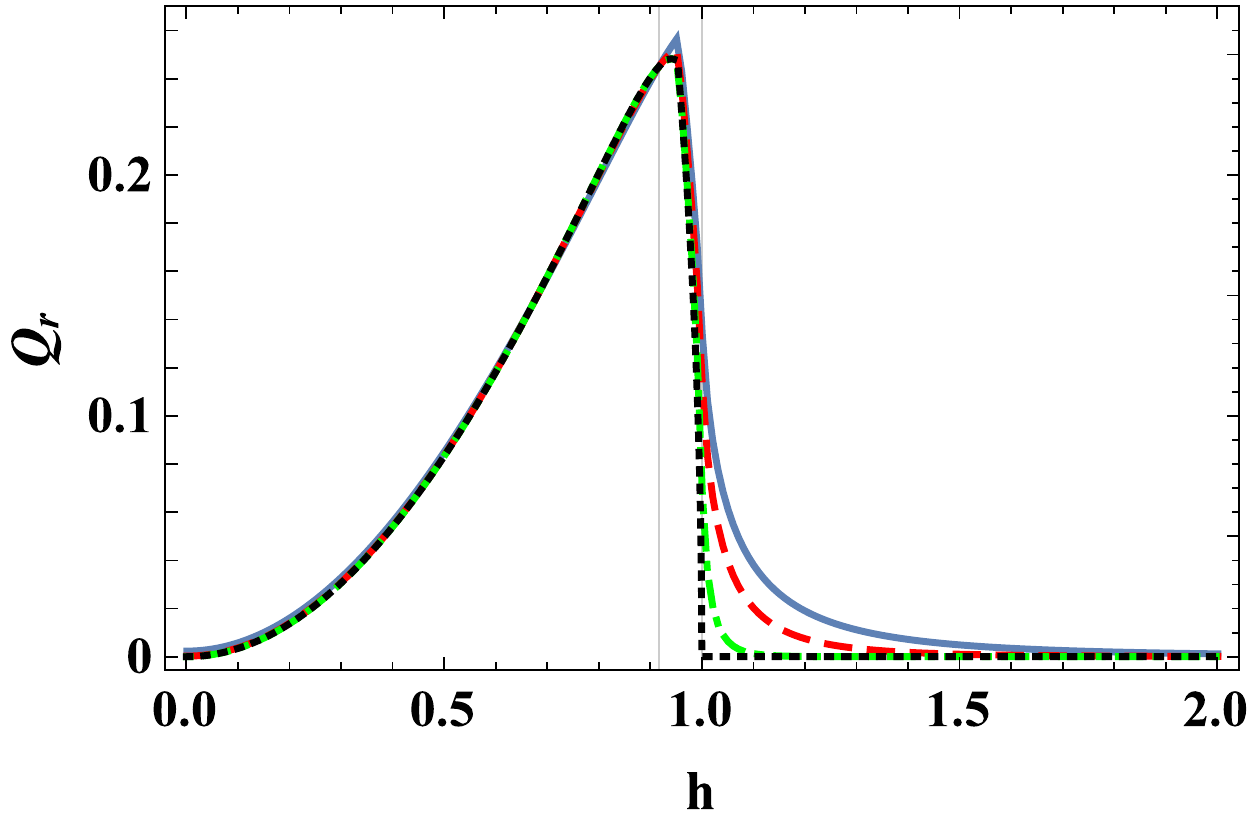}
	\includegraphics[width=7.65cm]{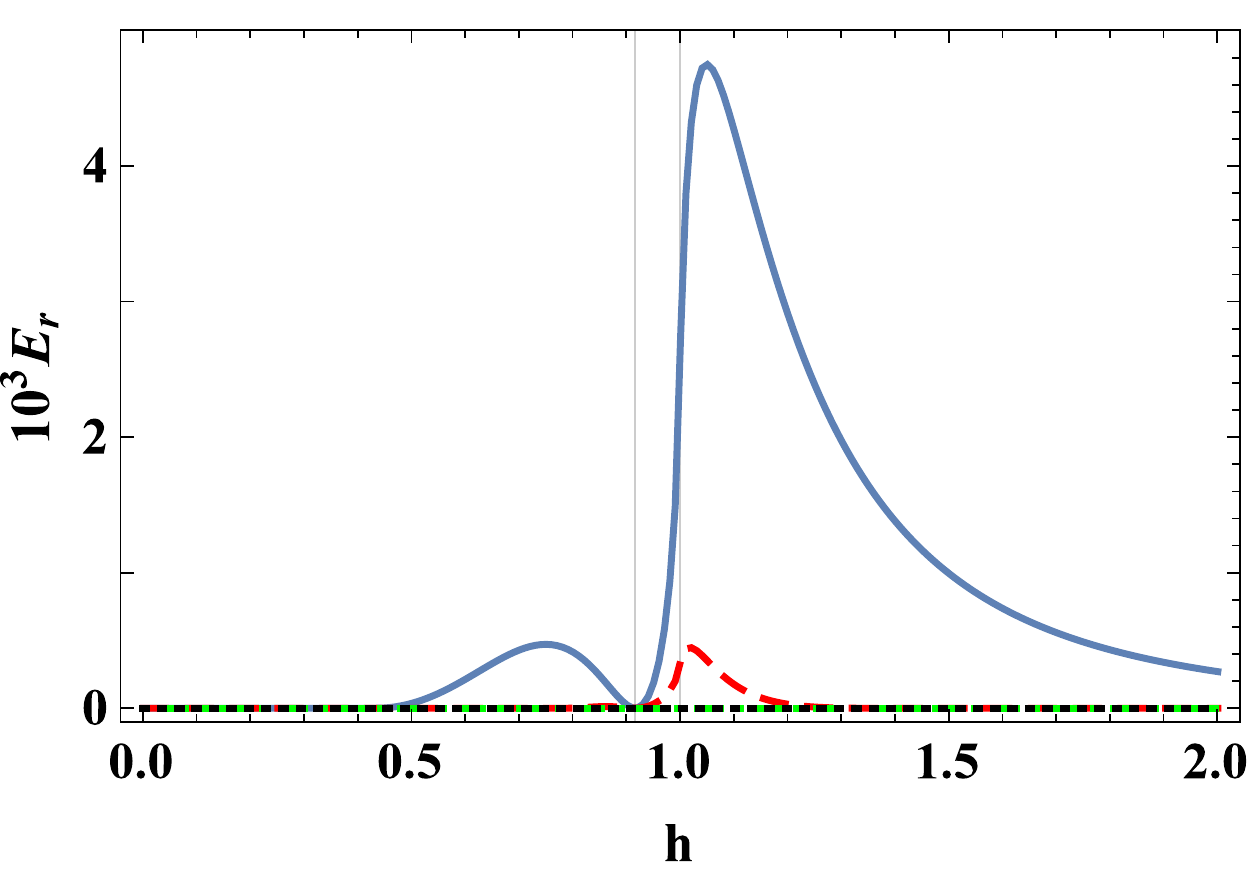}
	\caption{Two-body trace distance-based discord of response (left panel) and two-body trace distance-based entanglement of response (right panel) for symmetry-preserving ground states, in the thermodynamic limit, as functions of the external field $h$, in the case of $\gamma=0.4$, for different inter-spin distances $r$. Solid blue curve: $r=2$; dashed red curve: $r=3$; dot-dashed green curve: $r=8$; and dotted black curve: $r=\infty$. In both panels, the two solid vertical lines correspond, respectively, to the factorizing field (left) and to the critical field (right)}
	\label{fig:symnquantandentversushatvariousr}
\end{figure}

When increasing the inter-spin distance $r$, the pairwise entanglement of response $E_r$ and discord of response $Q_r$ behave even more differently (see Fig.~\ref{fig:symnquantandentversushatvariousr}). Due to the monogamy of the squared concurrence~\cite{Coffman2000,Osborne2006}, $E_r$ dramatically drops to zero as $r$ increases, except in a small region
around the factorizing field $h=h_f$ that gets smaller and smaller as $r$ increases, in agreement with the findings of Ref.~\cite{Amico2006}. On the other hand, while in the disordered and critical phases $Q_r$ vanishes as $r$ increases, in the ordered phase $Q_r$ survives even in the limit of
infinite $r$. Indeed, in both the disordered and critical phases, and when $r$ goes to infinity, the only nonvanishing one-body and two-body correlation functions in the symmetry-preserving ground states are
$\langle \sigma_i^z \rangle$ and $\langle \sigma_i^z \sigma_{i+r}^z\rangle$, so that the two-body reduced state can be written as
a classical mixture of eigenvectors of $\sigma_i^z \sigma_{i+r}^z$. On the other hand, in the ordered phase, also the two-body correlation function $\langle \sigma_i^x \sigma_{i+r}^x\rangle$ appears, while $\langle \sigma_i^x \rangle$ vanishes due to symmetry preservation, thus preventing the two-body marginal of the symmetry-preserving ground state from being a mixture of classical states.

\begin{figure}
	\includegraphics[width=8cm]{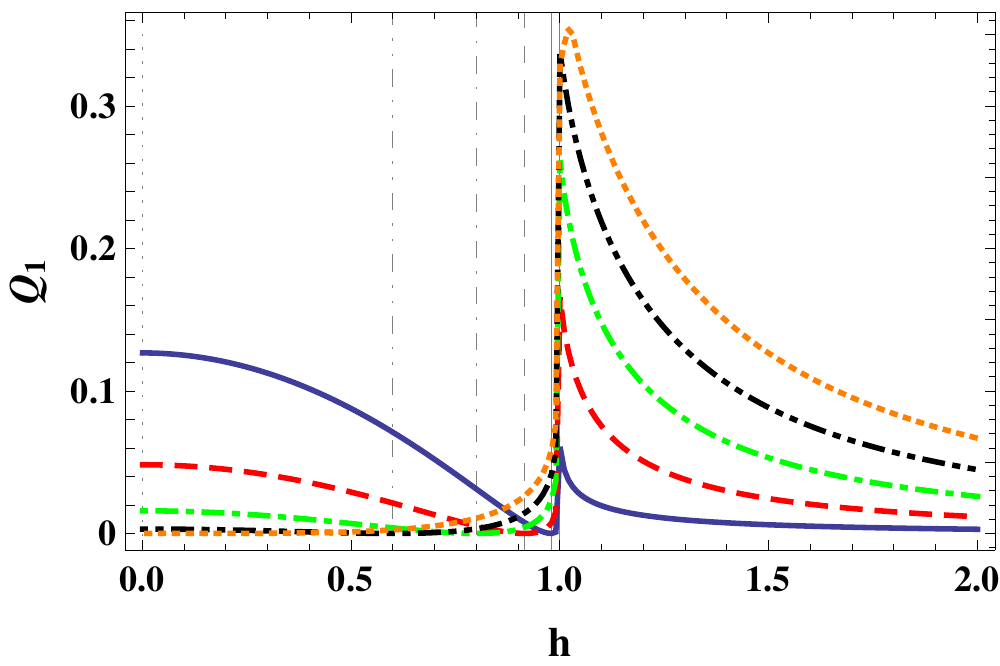}
	\includegraphics[width=8cm]{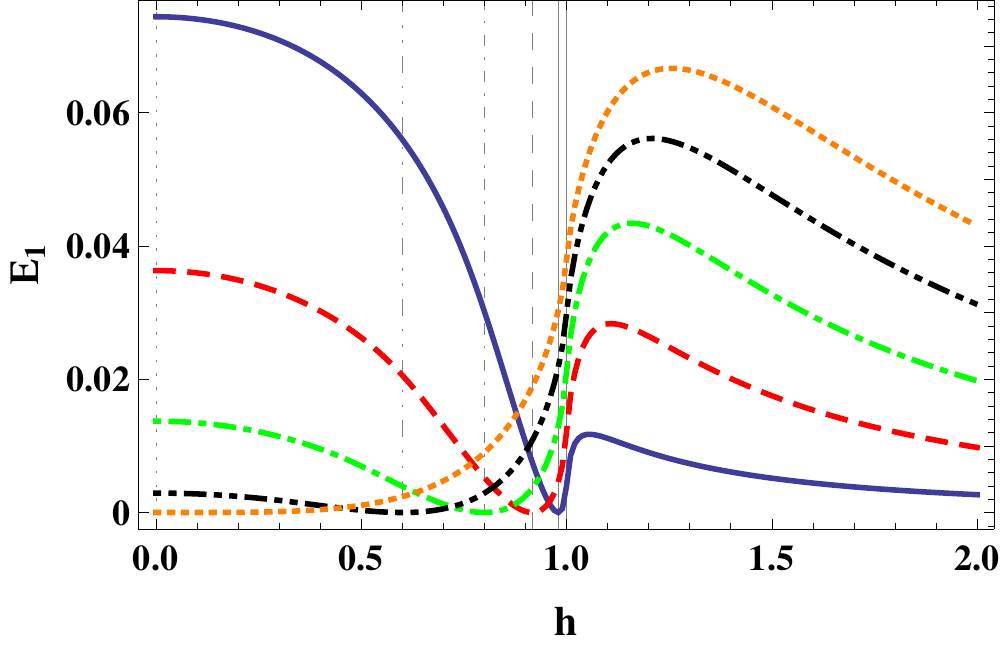}
	\caption{Nearest-neighbor trace distance-based discord of response (left panel) and nearest-neighbor trace distance-based entanglement of response (right panel) in MSBGSs as  functions of the external field $h$, for different values of the anisotropy $\gamma$. Solid blue curve: $\gamma=0.2$; dashed red curve: $\gamma=0.4$;
		dot-dashed green curve: $\gamma=0.6$; double-dot-dashed black curve: $\gamma=0.8$; and dotted orange curve: $\gamma=1$. In both panels, to each of these curves, there corresponds a vertical line denoting the associated factorizing field $h_f$. The rightmost vertical line denotes the critical point}
	\label{fig:symbrokennearestquantandent}
\end{figure}

\begin{figure}
	\includegraphics[width=8cm]{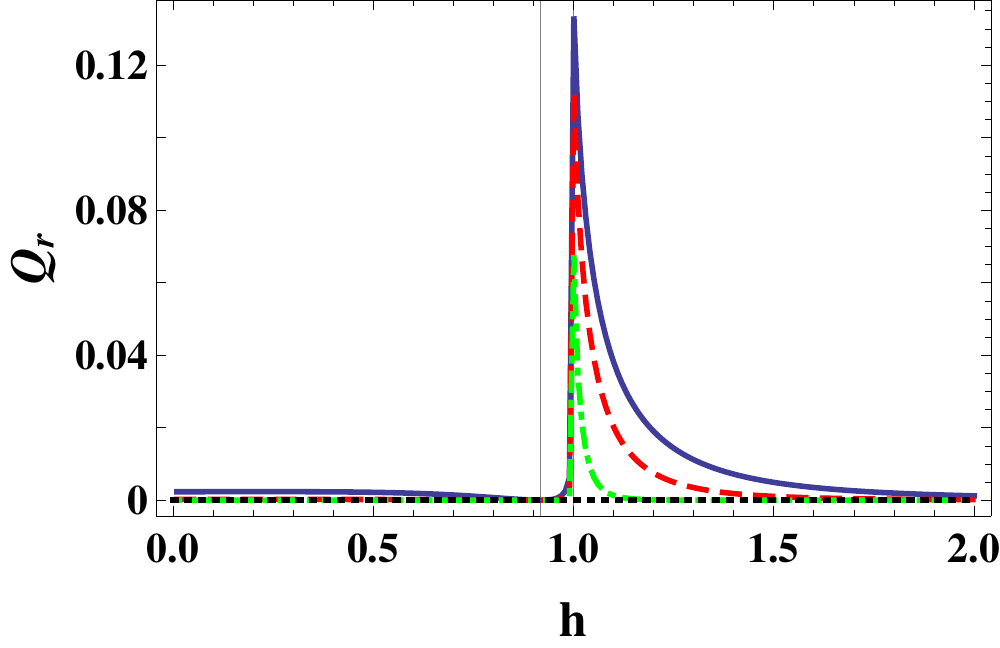}
	\includegraphics[width=8cm]{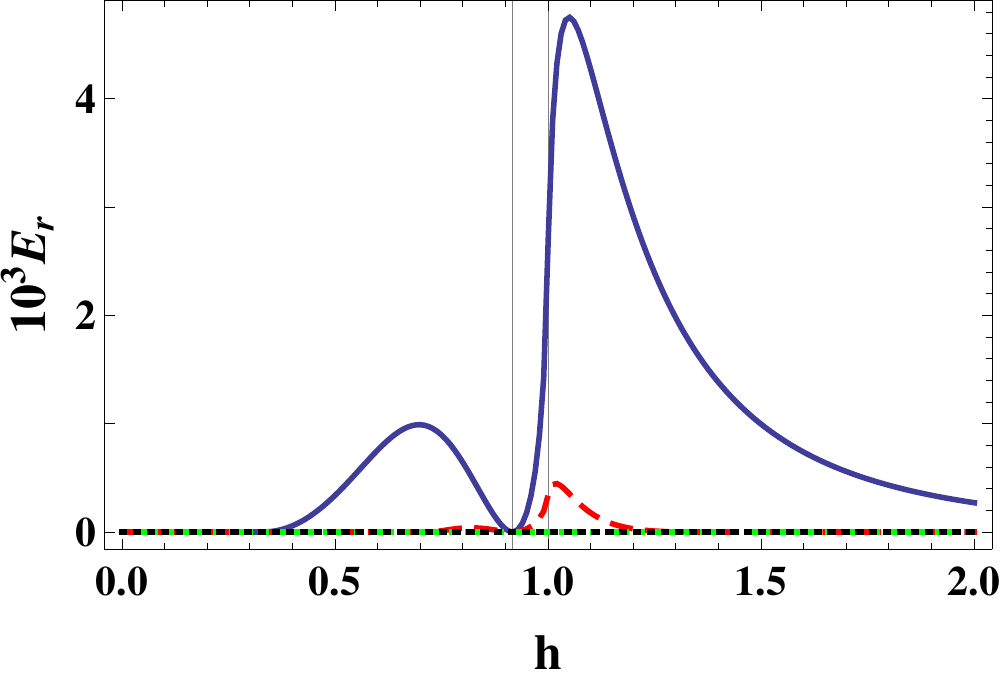}
	\caption{Two-body trace distance-based discord of response (left panel) and two-body trace distance-based entanglement of response (right panel) in MSBGSs as functions of the external field $h$, at $\gamma=0.4$, for different inter-spin distances $r$. Solid blue curve: $r=2$; dashed red curve: $r=3$;
		dot-dashed green curve: $r=8$; and dotted black curve: $r=\infty$. In both panels, the two solid vertical lines correspond, respectively, to the factorizing field (left) and to the critical field (right)}
	\label{fig:symbrokennquantandentversushatvariousr}
\end{figure}

\paragraph{Maximally Symmetry-Breaking Ground States}
\label{sec:symmmetrybrokengroundstate}

In this section, we move the focus of the comparison between two-body entanglement of response and discord of response from symmetry-preserving to MSBGSs. Spontaneous symmetry breaking manifests itself in the thermodynamic limit, in the ordered phase $h<h_c=1$ and for any nonzero anisotropy $\gamma$, so that hereafter we will restrict the region of the phase space under investigation accordingly.

Figure \ref{fig:symbrokennearestquantandent} shows that, as soon as symmetry breaking is taken into account,
only the discord of response is affected by symmetry breaking at the critical point $h_c=1$. In fact, according to Ref.~\cite{Osterloh2006}, the concurrence and, consequently, the two-body entanglement of response attain the same value for any $h\geq h_f$ both in the symmetry-preserving and MSBGSs. Otherwise, if $h<h_f$, there is a slight enhancement in the pairwise entanglement of response in the MSBGSs compared to the corresponding symmetry-preserving ones. Conversely, in general, the pairwise discord of response undergoes a dramatic suppression in the entire ordered phase $h<h_c$ when moving from symmetry-preserving to MSBGSs.

Considering the dependence on the inter-spin distance, we observe that the pairwise discord of response loses its long-range nature when moving from symmetry-preserving to MSBGSs (see Fig.~\ref{fig:symbrokennquantandentversushatvariousr}). More precisely, both the pairwise entanglement of response and the pairwise discord of response vanish asymptotically with increasing inter-spin distance. In the case of the pairwise entanglement of response, this result is again due to the monogamy of the squared concurrence~\cite{Coffman2000,Osborne2006}. In the case of the pairwise discord of response, it is instead due to the fact that not only the correlation function $\langle \sigma_i^x \sigma_{i+r}^x\rangle$ but also $\langle \sigma_i^x\rangle$ and $\langle \sigma_i^x \sigma_{i+r}^z\rangle$ are nonvanishing in the limit of infinite inter-spin distance $r$. This feature allows writing any two-spin reduced density matrix obtained from the MSBGSs as a classical mixture of eigenvectors of $O_i O_{i+r}$,
where $O_i$ is an Hermitian operator defined on the $i$th site as $O_i= \cos \beta \sigma_i^z + \sin \beta \sigma_i^x$ with $\tan \beta= \frac{\langle \sigma_i^x\rangle}{\langle \sigma_i^z\rangle}$.

Overall, the quantum correlations between any two spins decrease significantly in the entire ordered phase when symmetry breaking is taken into account and are almost entirely made up by contributions due to entanglement. In particular, at the factorizing field $h_f$, both the entanglement of response and the discord of response vanish. Indeed, we recall that the factorizing field $h_f$ owes its name to the two MSBGSs that are completely separable (product) at such value of the external magnetic field.

\subsection{Global Properties: Local Convertibility and Many-Body Entanglement Sharing}
\label{sec:symmetrybreakingorigin}

\begin{figure}
	\centering
	\includegraphics[width=9cm]{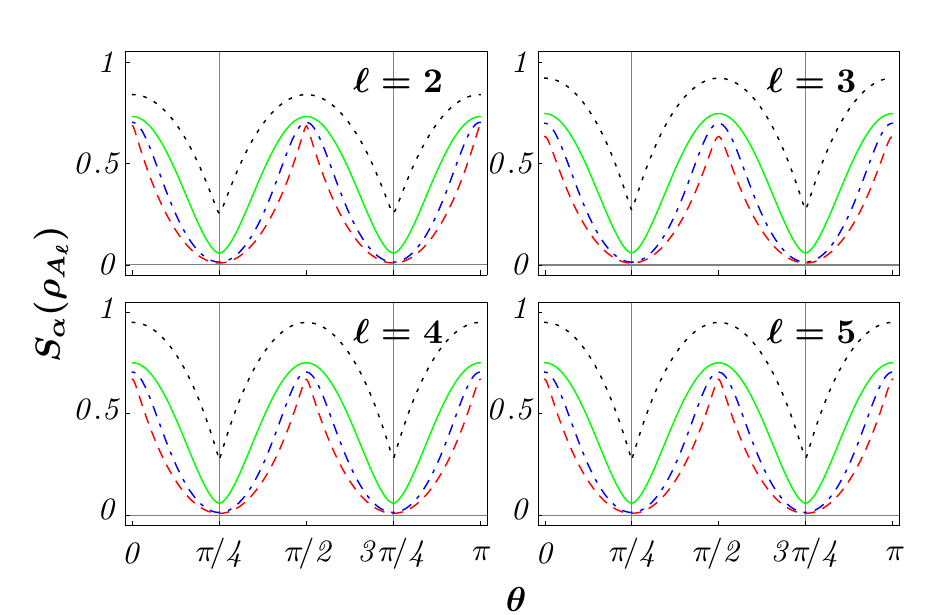}
	\caption{Behavior of the R\'enyi entropies $S_\alpha(\rho_A)$ as functions of the different ground states in the ordered phase, $h < h_c$,
		in the case of a subsystem $A_{\ell}$ made of $\ell$ contiguous spins. Each line stands for a different value of $\alpha$. Black dotted
		line: $\alpha=0.5$. Green solid line: $\alpha\rightarrow 1^+$ (von Neumann entropy). Blue dot-dashed line: $\alpha=3$. Red dashed line:
		$\alpha\rightarrow \infty$. The different ground states are parameterized by the superposition amplitudes $u=\cos(\theta)$ and $v=\sin(\theta)$. The two vertical lines correspond to the two MSBGSs, respectively obtained for $\theta=\pi/4$ and $\theta=3 \pi/4$. The Hamiltonian parameters are set at the intermediate values $\gamma=0.5$ and $h=0.5$. Analogous behaviors are observed for different values of the anisotropy and external field}
	\label{convertibilityversusell}
\end{figure}

We now investigate the nature of quantum ground states in the ordered phase concerning the properties of local convertibility of the global ground states and the many-body entanglement distribution.

\paragraph{Local Convertibility of Many-Body Quantum Ground States}

\begin{figure}
	\centering
	\includegraphics[width=9cm]{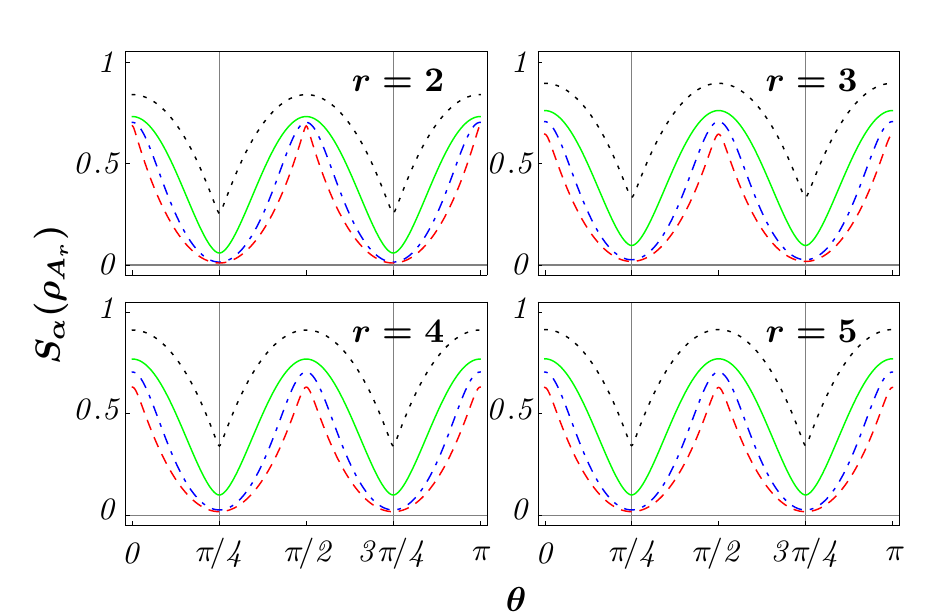}
	\caption{Behavior of the R\'enyi entropies $S_\alpha(\rho_A)$ as functions of the different ground states in the ordered phase, $h < h_c$,
		in the case of a subsystem $A_{r}$ made by two spins, for different inter-spin distances $r$. Each line stands for a different value of $\alpha$. Black dotted
		line: $\alpha=0.5$. Green solid line: $\alpha\rightarrow 1^+$ (von Neumann entropy). Blue dot-dashed line: $\alpha=3$. Red dashed line:
		$\alpha\rightarrow \infty$. The different ground states are parameterized by the superposition amplitudes $u=\cos(\theta)$ and $v=\sin(\theta)$. The two vertical lines correspond to the two MSBGSs, respectively obtained for $\theta=\pi/4$ and $\theta=3 \pi/4$. The Hamiltonian parameters are set at the intermediate values $\gamma=0.5$ and $h=0.5$. Analogous behaviors are observed for different values of the anisotropy and external field}
	\label{convertibilityversusr}
\end{figure}

We begin by studying the property of the local convertibility of quantum ground states in an ordered phase. 
It was previously shown that symmetric ground states are always locally convertible among themselves for $h_f < h < h_c$ and never for $h < h_f < h_c$~\cite{GiampaoloMontangero2013}. Here, thanks to the general methods developed before, we can investigate the local convertibility property of {\em all} quantum ground states in the ordered phase. In Fig.~\ref{convertibilityversusell}, we report the behavior of the R\'enyi entropies $S_\alpha$ as functions of the different ground states for a bipartition of the system in which subsystem $A$ is made of $\ell$ contiguous spins, while in Fig.~\ref{convertibilityversusr} we report it for subsystem $A$ made of two spins with various inter-spin distances.

We observe that the MSBGSs are the ground states characterized by the smallest value of all R\'enyi entropies, independently of the size $\ell$ of the subsystem and the inter-spin distance $r$. Therefore, all elements in the ground space are always locally convertible to a MSBGS, while the opposite is impossible. This first quantitative criterion for classicality is thus satisfied only by MSBGSs.

\subsection{Many-Body Entanglement Distribution}

We now compare symmetry-breaking and symmetry-preserving ground states with respect to entanglement distribution. The monogamy inequality quantifies in a simple and direct way the limits that are imposed on how bipartite entanglement may be shared among many parties~\cite{Coffman2000,Osborne2006}. For a given many-body system of $N$ $1/2$-spins, it reads:
\begin{equation}
	\tau (i|N-1) \geq \sum_{j \neq i} \tau(i|j) \; \; \; \; , \; \; \; \forall \; i \; .
	\label{monogamy}
\end{equation}
In the above expression, $\tau = C^{2}$ is known as the tangle, where $C$ is the concurrence~\cite{Hill1997,Wootters1998}, and the sum in the r.h.s. runs over all $N-1$ spins excluding spin $i$. The l.h.s. quantifies the bipartite entanglement between one particular, arbitrarily chosen, spin in the collection (reference spin $i$) and all the remaining $N-1$ spins. The r.h.s. is the sum of all the pairwise entanglements between the reference spin and each of the remaining $N-1$ spins. The inequality implies that entanglement cannot be freely distributed among multiple quantum parties $N \geq 3$, a constraint of quantum origin with no classical counterpart.

The residual tangle $\tilde{\tau}$ is the positive semi-definite difference between the l.h.s and the r.h.s in Eq.~(\ref{monogamy}). It measures the amount of entanglement not quantifiable as elementary bipartite spin--spin entanglement. Its minimum value compatible with monogamy provides yet another quantitative criterion for classicality.

Specializing, for simplicity but without loss of generality, to translationally invariant $XY$ spin systems in magnetically ordered phases, since the expectation value of $\sigma_i^y$ vanishes on every element of the ground space, the expressions of the tangle $\tau$ and the residual tangle $\tilde{\tau}$ for any arbitrarily chosen spin in the chain read, respectively,
\begin{eqnarray}
	\tau & = & 1- m_z^2-(u^*v+v^*u)^2 m_x^2 \; , \label{tangle} \\
	\tilde{\tau} & = & \tau - 2 \sum_{r=1}^{\infty} C_{r}^2(u,v) \geq 0 \label{residualtangle} \; ,
\end{eqnarray}
where $m_z = \langle e| \sigma_i^z | e \rangle = \langle o| \sigma_i^z | o \rangle$ is the on-site magnetization along $z$, the order parameter
$m_x = \langle e| \sigma_i^x | o \rangle =  \sqrt{\lim\limits_{r \to \infty} \langle e| \sigma_i^x \sigma_{i+r}^x| e \rangle}$, and
$C_{r}(u,v)$ stands for the concurrence between two spins at a distance $r$ when the system is in any one of the possible ground states $|g(u,v)\rangle$,
Eq.~(\ref{eq:groundstates}).

As already mentioned, by comparing the symmetric ground states with the MSBGSs, the spin--spin concurrence is larger in the MSBGSs~\cite{Osterloh2006} if $h < h_f < h_c$, where $h_f = \sqrt{1 - \gamma^2}$ is the factorizing field, while for $h_f < h < h_c$ they are equal. We have verified that these two results are much more general. We have compared all ground states (symmetric, partially symmetry breaking, and MSBGSs), and we have found that for $h < h_f < h_c$ the spin--spin concurrences are maximum in the MSBGSs for all values of the inter-spin distance $r$, while for $h_f < h < h_c$ and for all values of $r$ they are independent of the superposition amplitudes $u$ and $v$ and thus acquire the same value irrespective of the chosen ground state. 

Finally, it is immediate to see that the third term in the r.h.s. of Eq.~(\ref{tangle}) is maximized by the two MSBGSs. Collecting all these results, it follows that the many-body, macroscopic multipartite entanglement, as quantified by the residual tangle, is minimized by the two MSBGSs and therefore also this second quantitative criterion for classicality is satisfied only by the MSBGSs among all possible quantum ground states.

\section{Conclusions}
\label{conclusionSec}

We have shown that phases characterized by topological order or systems prepared in a ground state supporting edge states lack differential global convertibility, due to the long-range entanglement that these conditions entail. Moreover, the breaking of dLC is detectable even more clearly when small partitions are considered. This means that dLC constitutes a semi-local probe to detect LRE, which is instead an inherently nonlocal property, usually accessible through string operators stretching for distances much larger than the correlation length $\xi$.

We also argue that the competition between $\xi$ and LRE is the reason for the lack of dLC, because when the size of a partition becomes comparable with $\xi$, local correlations reduce the LRE between the farthest point of the partition. Thus, as $\xi$ increases, bulk entanglement increases, but LRE decreases, thus creating a non-monotonous behavior in the R\`enyi entropies as $\alpha$ is varied.

Since LRE is an intrinsic property of a quantum phase that cannot be created or destroyed, except by passing through a phase transition, our analysis highlighted once more the higher computational power phase with LRE is endowed. The lack of dLC renders them {\it more quantum} that phases that are locally convertible.

This intuition was then used to investigate the classical nature of globally ordered phases associated with nonvanishing local order parameters and spontaneous symmetry breaking. We have put on quantitative grounds the long-standing conjecture that the maximally symmetry-breaking ground states (MSBGSs) are macroscopically the most classical ones among all possible ground states. We have proved the conjecture by introducing and verifying two independent quantitative criteria of macroscopic classicality. The first criterion states that all global ground states in the thermodynamic limit are locally convertible to MSBGSs, while the opposite is impossible. The second criterion states that the MSBGSs are the ones that satisfy at its minimum the monogamy inequality for globally shared bipartite entanglement and thus minimize the macroscopic multipartite entanglement as quantified by the residual tangle. We have thus verified that, according to these two criteria, the MSBGSs are indeed the most classical ones among all possible quantum ground states.

\section*{Acknowledgments}

We thanks our collaborators with whom the results presented in this chapter have been originally published, in particular Jian Cui, Heng Fan, Mile Gu, Alioscia Hamma, Vlatko Vedral, Lukasz Cincio, Siddhartha Santra, Paolo Zanardi, Marco Cianciaruso, Leonardo Ferro, Giuseppe Zonzo, Wojciech Roga, Massimo Blasone, and Fabrizio Illuminati.
VK is supported by the  U.S. Department of Energy, Office of Science, National Quantum Information Science Research Centers, Co-design Center for Quantum Advantage (C2QA) under contract number DE-SC0012704.
SMG and FF acknowledge support from the QuantiXLie Center of Excellence, a project co-financed by the Croatian Government and European Union through the European Regional Development Fund -- the Competitiveness and Cohesion (Grant KK.01.1.1.01.0004) and from the Croatian Science Foundation (HrZZ) Projects No.  IP--2019--4--3321.


\begin{thebibliography}{99}
\bibitem{caveat} For moderate $\xi$, also S2 splits in the right way, and it is more convenient to measure \cite{abanin, measurerenyi}
\end{thebibliography}
\end{document}